%% file: 00_main.tex
\documentclass[sigconf]{acmart}

\usepackage{listings}
\usepackage{wrapfig}
\usepackage{array}
\usepackage{multirow}
\usepackage{blindtext}
\usepackage{url}
\usepackage{hyperref}
\usepackage{enumitem}

\copyrightyear{2025}
\acmYear{2025}
\setcopyright{cc}
\setcctype{by-nc}
\acmConference[UIST '25]{The 38th Annual ACM Symposium on User Interface Software and Technology}{September 28-October 1, 2025}{Busan, Republic of Korea}
\acmBooktitle{The 38th Annual ACM Symposium on User Interface Software and Technology (UIST '25), September 28-October 1, 2025, Busan, Republic of Korea}\acmDOI{10.1145/3746059.3747605}
\acmISBN{979-8-4007-2037-6/2025/09}

\usepackage{bbding}%
\AtBeginDocument{%
  \providecommand\BibTeX{{%
    \normalfont B\kern-0.5em{\scshape i\kern-0.25em b}\kern-0.8em\TeX}}}


\begin{document}

\title{Computational Scaffolding of Composition, Value, and Color for Disciplined Drawing}

\newcommand{\commentText}[1]{#1}    
\newcommand{\TODO}[1]{\commentText{{\color{red}[\textbf{\textsc{TODO}}: \textit{#1}]}}}
\newcommand{\jiaju}[1]{\commentText{{\color{blue}[\textbf{\textsc{JM}}: \textit{#1}]}}}
\newcommand{\jingyi}[1]{\commentText{{\color{teal}[\textbf{\textsc{JL}}: \textit{#1}]}}}
\newcommand{\chauvu}[1]{\commentText{{\color{purple}[\textbf{\textsc{CV}}: \textit{#1}]}}}
\newcommand{\asya}[1]{\commentText{{\color{orange}[\textbf{\textsc{AL}}: \textit{#1}]}}}
\newcommand{\camready}[1]{\commentText{{{#1}}}}

\newcommand{\EDITED}[1]{{#1}}
\newcommand{\changes}[1]{{#1}} 

\renewcommand{\changes}[1]{{#1}}

\newcommand{\systemName}[1]{ArtKrit}

\author{Jiaju Ma}
\affiliation{
  \institution{Stanford University}
  \city{Stanford}
    \state{California}
  \country{USA}}

\author{Chau Vu}
\authornote{Both authors contributed equally to this research.}
\affiliation{
  \institution{Pomona College}
  \city{Claremont}
  \state{California}
  \country{USA}
}
 
\author{Asya Lyubavina}
\authornotemark[1]
\affiliation{
  \institution{Pomona College}
  \city{Claremont}
  \state{California}
  \country{USA}
}

\author{Catherine Liu}
\affiliation{
  \institution{Claremont McKenna College}
  \city{Claremont}
    \state{California}
  \country{USA}
}

\author{Jingyi Li}
\affiliation{%
  \institution{Pomona College}
  \city{Claremont}
  \state{California}
  \country{USA}}

\begin{abstract}
One way illustrators engage in disciplined drawing---the process of drawing to improve technical skills---is through studying and replicating reference images. 
However, for many novice and intermediate digital artists, knowing how to approach studying a reference image can be challenging.
It can also be difficult to receive immediate feedback on their works-in-progress.
To help these users develop their professional vision, we propose \systemName{}, a tool that scaffolds the process of replicating a reference image into three main steps: composition, value, and color. 
At each step, our tool offers computational guidance, such as adaptive composition line generation, and automatic feedback, such as value and color accuracy. 
Evaluating this tool with intermediate digital artists revealed that \systemName{} could flexibly accommodate their unique workflows and encourage reflection-in-action on their drawing process. As a design probe, \systemName{} suggests that computational scaffolds \changes{that enact new norms} may drive new artistic insights.

\end{abstract}


\begin{teaserfigure}
  \includegraphics[width=\textwidth]{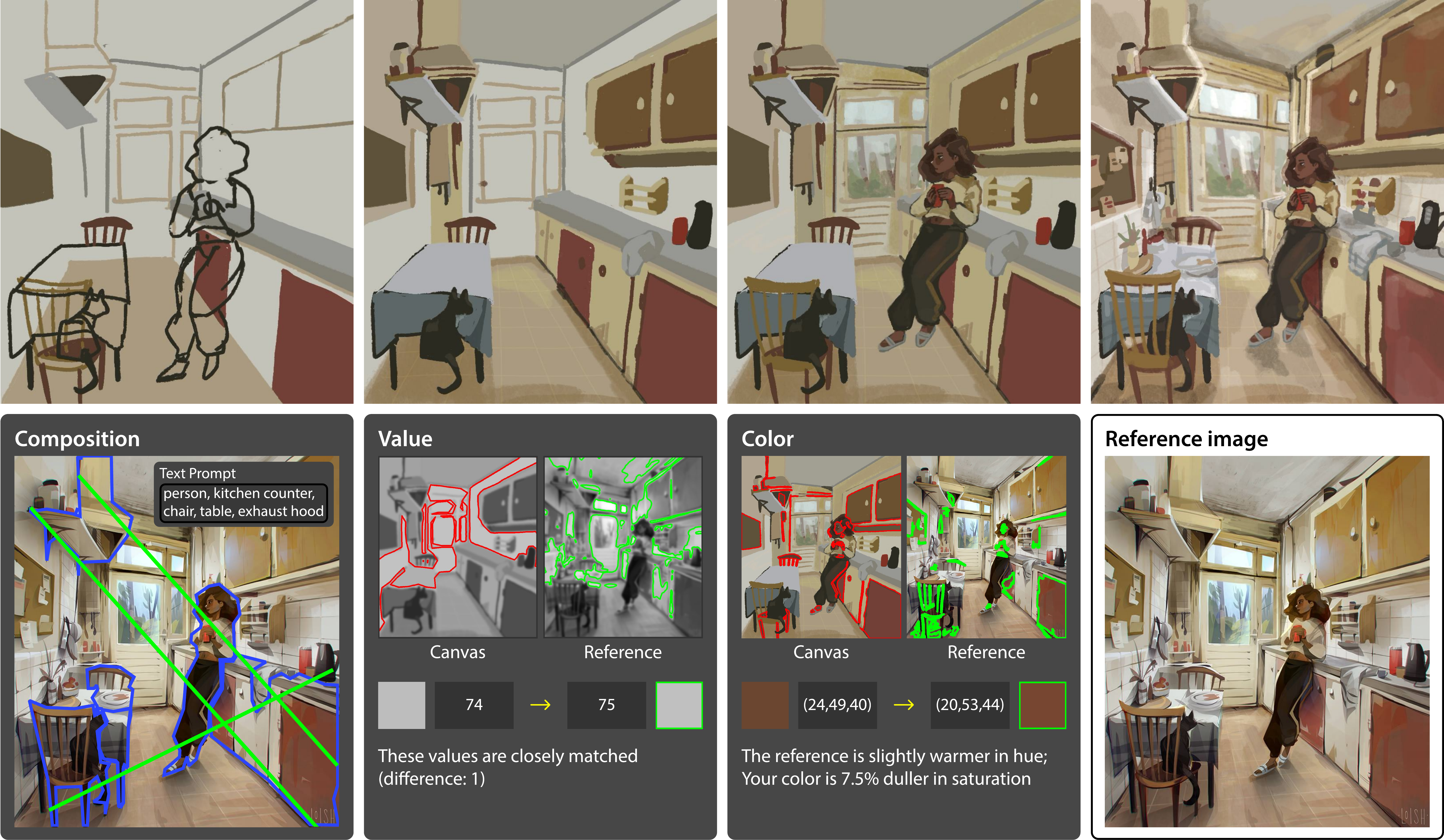}
  \caption{%
  We propose \systemName{}, a tool that scaffolds the digital drawing process of replicating a reference image into three steps: composition, value, and color.
  Top row: a user's drawing process. Bottom row: computational guidance and feedback that our system provides for each step. 
  We provide object-based composition lines to help with spatial positioning, and visualize differences in value and color with verbal suggestions.
  \textit{Reference image: interior practice // kitchen by Loish (2021, digital).}
  }
  \label{fig:teaser}
\end{teaserfigure}

\maketitle

\input{01_introduction}
\input{02_related_work}
\input{03_design}
\input{04_tool}
\input{05_method}
\input{06_eval}

\input{07_discussion}
\input{08_conclusion}

\section{Acknowledgments}
The authors would like to extend a huge thanks to the artists who participated in evaluating this work. Gratitude also to EK Kim and Vivian Wang who worked on designing earlier iterations of this tool, Kate Nikles and Jacob Ritchie for their feedback on the paper, Chuan Yan and Andy Xu for their insights on the technical aspects, and our anonymous reviewers for helping us tighten our arguments.

\bibliographystyle{ACM-Reference-Format}
\bibliography{10_reference}

\end{document}

%% file: 01_introduction.tex
\section{Introduction}

Computational drawing tools, such as Procreate~\cite{procreate}, Photoshop~\cite{photoshop}, and Krita~\cite{krita}, have become widespread for digital art. Just like traditional artists, digital artists have to practice drawing in their medium and develop proficiency in their tools. We explore how to computationally support this process of practicing digital art---what is sometimes termed as ``disciplined drawing.'' Artists often engage in disciplined drawing by studying and replicating reference images, such as paintings by more skilled ``master'' artists~\cite{drawing-discipline}. The goal of disciplined drawing is to practice technical execution skills, such as accurately matching the shape language or colors of the reference image. This goal is in contrast to making art as a response to culture or to communicate feelings \cite{ai-art-artists}; instead, artists often do disciplined drawing as daily warm-ups or in preparation for a larger piece \cite{hale1989drawing}. 

For many developing artists, learning how to have the professional vision \cite{goodwin-professional, color-field} to break down these reference images into concrete drawing steps can be a challenge. For instance, in figurative drawing, artists need to know how 3D forms can be flattened into 2D, how shadow and light bounce from these forms, how the forms can be composed in a visually pleasing or meaningful way, and how colors can harmonize together \cite{hale1989drawing}. These parts are difficult to disentangle and sense-make from looking at a finished, static reference image.

For these reasons, the Western tradition of drawing teaches artists how to break down images---namely into different drawing steps of composition, value, and color (please refer to Section \ref{sec:design:process} for a more detailed review). For example, by focusing their attention on the composition and shape language of a piece, artists choose not to ``look'' at the shadows and hues until later. Drawing is about looking and consciously abstracting details into personalized mark making \cite{petherbridge1991primacy}. 
Artists draw from ``master studies'' not to create a compelling final artifact, but for the \textit{process}---through the process of studying another's work, artists can gain valuable insights and transferable knowledge into their own workflows and creative practices \cite{creative-version-contorl-sterman}. 

Thus, the aim of our work is not to get artists to perfectly reproduce reference images, but to better understand and reflect on their own digital drawing processes.
We present \systemName{}, a tool to support disciplined digital drawing through breaking down reference images into \textit{composition}, \textit{value}, and \textit{color} drawing steps. Each step offers automatically generated guidance as well as feedback grounded by comparing the user's in-progress canvas to the reference image (Figure~\ref{fig:teaser} and \ref{fig:usage_scenario}). 
The combination of our computational guides and contextual feedback supports illustrators in developing their professional vision \cite{goodwin-professional, color-field}, knowing ``how to look'' at both the reference and their own drawing, \changes{as well as encouraging reflection as a process aesthetic \cite{kreminski2021reflective}}.

We note that artists are known for having idiosyncratic processes \cite{learn-visual-artists-li} and that replicating ``master studies'' implies the reference images enforce norms around what is ``good'' art. \changes{Digital drawing workflows can take many different paths, and some digital artists may not even conform to the composition-value-color breakdown suggested by traditional Western drawing.} 
In line with a recent call for tools that mitigate the power imbalances between researchers and artists \cite{power-cst}, we frame \systemName{} as an \textit{artistic support tool}, a system flexible enough to fit multiple workflows, an extension that can be ignored as easily as it is used, and a software that focuses on delivering interactions useful to a creative practice rather than auto-generating results. 
\changes{We present \systemName{}'s guidance and feedback not as ``the right way to draw,'' but as showing what the reference images are like for users to see ``what the machine sees'' \cite{phraselette-10.1145/3715336.3735832}.}


The main technical contribution of \systemName{} is a new method to adaptively generate \textit{composition lines} based on user-specified objects of interest~\cite{arnheim1954art, mukherjee2019communicating}.
We capture user intent through both textual descriptions and on-canvas bounding boxes, and output lines describing spatial relations of the intended objects by employing a modified RANSAC algorithm on feature points produced by Grounded SAM~\cite{ren2024gsam} (Figure~\ref{fig:composition}).
We additionally provide value and color guidance.
For \textit{value}, we provide guidance through blurring and grayscaling the reference image to generate a value thumbnail to mimic peripheral vision for value perception~\cite{livingstone2002vision}.
For \textit{color}, we visualize regions of the same color isolated on a white background to help with accurate color matching~\cite{albers2006interaction}.
In terms of feedback, we generate polygon contours that show both silhouette and position for composition, as well as value and color ground truths from the reference image, which users can compare with their work-in-progress at any time. 
We offer feedback across multiple levels of abstraction (e.g., numbers, visuals, and natural language as shown in Figure~\ref{fig:teaser} and \ref{fig:usage_scenario}).
We develop \systemName{} as an extension for Krita~\cite{krita} (a free and open-source painting program),
\changes{available at \url{https://github.com/doodlelab/ArtKrit}}.

In evaluating \systemName{}, we not only wanted to collect qualitative feedback around our design decisions, but also treat it as a design probe \cite{mattelmaki2006design-probe} to generate insights for HCI researchers around supporting digital drawing workflows. Five intermediate digital artists used \systemName{} to create ``master studies'' of their choice; then, two professional art educators gave feedback on these drawings as a comparative baseline before seeing our system. Study participants commented \systemName{} supported their unique workflows; we observed instances of reflection-in-action \cite{schon2017reflective} where they reflected on both their artwork and process. The professional feedback largely matched with our tool's composition, value, and color breakdown, but also highlighted limitations we cannot currently computationally support, such as stroke quality, mark making, and interpretation of artistic intent and norms. Finally, we suggest that \changes{analytical} computational scaffolds that are purposefully ``brittle'' may drive new artistic insights and different ways of seeing.

\vspace{2mm}
\noindent
In summary, we make the following contributions:
\begin{itemize}[leftmargin=1em]
    \item \systemName{}, a tool that scaffolds the process of disciplined drawing through computational guidance and feedback for composition, value, and color.
    \item A novel algorithm for generating object-based composition lines.
    \item Results from using \systemName{} as a design probe, including design takeaways and a small but rich dataset of professional feedback on six ``master studies'' (please refer to the appendix).
\end{itemize}


\noindent
\textit{Note: All the reference artwork in this paper is either public domain, used with artist permission, or under fair use copyright policy.}





%% file: 02_related_work.tex
\section{Related Work}

While many available tools~\cite{qin2023isda} assist users with completing drawings (e.g., ControlNet~\cite{zhang2023controlnet} generates a full image from a text prompt and rough sketch), our work focuses on providing computational guidance to help users \textit{learn} to draw by themselves. Compared to domains like UI and graphic design where feedback follows more formal rules~\cite{ui-feedback-peitong}, learning how to draw ``well'' is more subjective.

\subsection{Computational drawing guidance}


Most prior work on facilitating learning to draw has focused either on a selected portion of the drawing process (e.g., block-in shapes~\cite{iarussi2013da} and shading~\cite{williform2019drawmyphoto}) or on specific drawing domains (e.g., portraits~\cite{dixon2010icandraw, xie2014ps, huang2022dualface} and 3D design~\cite{hennessey2017h2s, williford2017sketchtivity}).
For example, The Drawing Assistant~\cite{iarussi2013da} helps users more accurately draw shapes of individual objects in photos by providing contours lines and skeletons.
Sketch Sketch Revolution~\cite{fernquist2011ssr} represents the drawing process as a sequence of strokes and guides users through learning how to replicate each stroke.
Focusing on value, DrawMyPhoto~\cite{williform2019drawmyphoto} helps novices learn how to render realistic grayscale drawings from photos by guiding them through five auto-generated steps on shading

For specific domains,
iCANDraw~\cite{dixon2010icandraw} generates step-by-step instructions with the goal of accurate proportions, while PortraitSketch~\cite{xie2014ps} automatically beautifies user strokes to better match the reference image.
More recently, dualFace~\cite{huang2022dualface} provides both global guidance (relative positioning of facial features) and local guidance (contours of individual features).
%
How2Sketch~\cite{hennessey2017h2s} and SketchTivity~\cite{williford2017sketchtivity} work with 3D perspective drawings, a type of linework used mostly in fields like industrial design and architecture.
How2Sketch automatically generates drawing tutorials, while SketchTivity provides feedback on quality and speed of strokes.

In contrast to prior work,
\changes{our system guides users through the end-to-end drawing \textit{process},}
rather than focusing on generating correct artifacts.
Informed by prior work based on foundational textbooks \cite{rockman2009drawing, loomis1947creative} and the illustration curriculum in the West \cite{angelacademy2025art, florenceacademy2025art, grandatelier2025art}, we structure the drawing process into composition, value, and color and provide computational guidance and visual and verbal feedback for each of the three stages (Figure~\ref{fig:teaser} and \ref{fig:usage_scenario} and Table~\ref{tab:system}).

\subsection{Interacting with value and color}
Supporting users to better work with and understand color in drawings has been a focus of much HCI research. 
%
Histomages~\cite{chevalier2012histo} uses the HSV color representation and converts an input image into three histograms for each of hue, saturation, and value.
This allows the user to edit each histogram individually, such as shifting the value distribution.
Mondrian~\cite{shi2024mondrian} provides three different kinds of color palettes for an image: 1) a 1D list of top $N$ dominant colors, 2) the 1D list of colors with variable widths in proportional to their frequencies in the image, and 3) a 2D array of dominant colors spatially matching where they appear in the image.
Both Color Triads \cite{color-triads-10.1145/3386569.3392461} and Color Builder \cite{color-builder-10.1145/3290605.3300686} display image colors in a triangular arrangement to facilitate color changing via direct manipulation.
Playful Palette~\cite{shugrina2017pp} keeps a history of color palettes as the user completes and edits a drawing for learning and remixing purposes. 

While these works present new representations and abstractions to work with color,
we choose to present feedback in the HSV color space because of its high consistency with human color perception~\cite{chernov2015integer} and its separation of \textit{value} (V channel) from \textit{color} (H and S channels).
This allows us to provide independent guidance on value and color.
Our tool visualizes and compares the dominant values and colors of a user's canvas with the reference image so they can make manual adjustments to learn how values and colors might better match.

\subsection{Automatic composition analysis}
\label{sec:related:composition}
Composition refers to spatial relationships between various elements in a scene.
Prior work has focused on finding specific types of compositional structures in images.
Zhou et al.~\cite{zhou2017ddvp} use a contour detection-based method to find dominant vanishing points in natural scenes.
He et al.~\cite{he2018dtp} look for triangular constructions in portraits by fitting triangles to edges in an image.
E et al.~\cite{e2020composition} adopt the harmonic armature widely used in photography composition.
They detect salient objects as users move their camera around to highlight lines in the armature that overlap with these objects to provide composition guidance at capture time.
However, these works mostly operate on natural (i.e., photographic) images, which are quite different from digital drawings in terms of realism and stylization.
The specific focus on a small set of compositional constructs prevents these works from providing comprehensive and flexible guidance to artists.

Besides low-level well-defined compositional features, a body of research aims to detect ``semantic lines'' that characterize layout by separating different semantic regions of an input natural image~\cite{lee2017sld}.
Lee et al.~\cite{lee2017sld} first formulated this task and proposed a dataset of images with manually annotated semantic lines.
They then train a CNN-based model for line detection.
Zhao et al.~\cite{zhao2022hough} improved on this method by incorporating the classical Hough transform technique into a learning-based framework.
More recently, Ko et al.~\cite{ko2024slcd} proposed SLCD, a two-stage approach in which they find $K$ semantic line candidates with a ResNet-based model~\cite{he2016deep} and then find the best combinations of these candidates via a scoring function as the final output.
Similarly, Zhang et al.~\cite{zhang2023reconstructing} employ another two-stage design to detect leading lines, a specific type of semantic lines.
They extract line candidates via edge detection and then group them into leading lines based on proximity.
However, ``semantic lines'' in this type of work are not well-defined and are subject to interpretations of the creators of the annotated datasets, who might not often agree with how and where artists draw semantic lines \cite{cole2008people}.
Our tool draws composition lines to describe spatial relations between objects of interest and support users in directly defining what these objects are through a combination of textual descriptions and bounding boxes drawn on canvas (Figure~\ref{fig:composition}).


%% file: 03_design.tex
\section{Design Study}
\label{sec:design}
To offer design rationale for \systemName{}, we summarize the drawing process and report on the results of a small formative study conducted to test initial design ideas. 

\subsection{Background: the drawing process}
\label{sec:design:process}
We analyzed art books and course materials of famous Western academies to learn how drawing has been traditionally taught.
We found that professional artists and educators have traditionally scaffolded the drawing process around three core elements: \textit{composition}, \textit{value}, and \textit{color}. This process is fundamental in introductory to advanced art books \cite{rockman2009drawing, edwards1997drawing}. For example, Andrew Loomis’s Creative Illustration---a seminal art book for professional illustrators---introduces the foundations of drawing in its first three chapters: line (composition), tone (value), and color \cite{loomis1947creative}. 

The first element, \textit{composition}, involves arranging elements in a given space to achieve balance and harmony, both between individual components and in relation to the art work as a whole \cite{dow1914composition}. \textit{Value}, or tone, defines the range of light and shadow in a drawing and is often taught before color. The ability to understand and master value is essential for using \textit{color} in later stages, as it helps convey depth and texture of a drawing~\cite{rockman2009drawing, edwards1997drawing, loomis1947creative}.  Many classical art academies worldwide \cite{angelacademy2025art, florenceacademy2025art, grandatelier2025art} also reflect this hierarchical approach to teaching drawing. Following these programs, students first master tonal drawing—typically using graphite or charcoal in grayscale—before advancing to color painting in later years.

\subsection{Formative study method}

We built a Figma desktop prototype to test our initial design ideas around composition, value, and color scaffolds. We recruited three artists (A1, A2, A3) to do a master study on Procreate of a reference image they chose. We selected modern drawings with a variety of compositions: multiple people, an indoor scene, and an outdoor scene (please refer to the appendix for the formative study reference images and participant drawings).
As participants drew on an iPad, our prototype was open on a desktop.


Our composition scaffolds were various grids that we manually created for each reference art piece. We had simple grids (like rule of thirds), custom leading line annotations, and also traced the objects as ``ground truth'' composition feedback. Before the study, the experimenter loaded these grids as hidden layers on the Procreate canvas; the participants could see the grids overlaid on the reference image on our desktop prototype. The value scaffold showed variations of blurred, grayscale representations of the reference image, as well as visualizations that isolated light tones, midtones, and shadows. 
In the color section, we displayed dominant colors manually extracted from the reference image. During the study, the experimenter would input a participant's chosen colors into a custom-built Figma color comparison plugin that output the \textit{Delta E} color difference metric~\cite{delta-e}.

%
%
All participants had previous experience with both traditional and digital art. 
Each artist self-identified their skill level on a scale ranging from novice to professional (A1: intermediate-advanced, A2: novice-intermediate, A3: intermediate).
Each study session lasted about one hour, and began with a short introduction and walkthrough of our prototype.
The participants then took 50 minutes to replicate their reference image, while being prompted to think aloud and explain their decisions to use (or not use) certain scaffolds. After the drawing phase, we conducted a short post-interview on how the prototype affected their process and what features they found most helpful.


\subsection{Formative study findings}
\label{sec:design:findings}
We summarize our findings into four main themes that guided our design iteration of \systemName{}.

\subsubsection{Subjective preferences for composition guidance}
Different participants found different kinds of composition grids helpful.
A1 mentioned that they liked to use the 2-3 grid that split the reference into six even rectangles, as it divided the image into smaller areas to focus on and made it easier to position objects.
On the other hand, A2 indicated that their favorite scaffold was the leading lines that emphasized a vanishing point.
They liked its familiarity with their existing practice, ``I remember seeing something like this in a middle school classroom where it showed you how to draw depth between objects.''
Overall, we observed there was no one-size-fits-all solution for providing guidance on composition. It depended on factors such as reference image content, stage in the drawing process, and users' backgrounds and histories.

\subsubsection{Over-reliance on composition guidance}
While all participants found the composition grids to be helpful, we found that overlaying ``ground truth'' object contours on the canvas led less experienced participants to directly trace over them, as the case with A2 tracing the outlines of the buildings in their reference.
This case demonstrated that we needed to carefully design our guidance and feedback system to facilitate the \textit{practice} of drawing skills, not the \textit{completion} of a drawing.


\subsubsection{Natural language and color feedback}
Our prototype displayed numeric Delta E values (the smaller the value, the closer the colors) as feedback for color accuracy. 
Both A1 and A2 noted that they would have preferred natural language feedback about their choices instead of numeric feedback, as it was difficult to interpret what the numbers meant. A1 added that seeing the hex values of ground truth colors was not helpful in determining how to improve their color choices, and instead suggested receiving feedback like ``this is too saturated or this is too blue by something percent.'' A3 mentioned the different color and brightness settings between the desktop computer and iPad made it hard to color match across displays, which encouraged us to build a tool that would show feedback on the same display as the canvas. 


\subsubsection{Varying workflows}
All participants directly skipped to color after composition; both A1 and A2 ignored the value feature, while A3 created a separate value drawing after their color drawing because they had extra time. 
A2 explained that they did not find value guidance helpful as they did not draw in grayscale. Although the Delta E values were confusing, A2 appreciated how our color scaffolds visually paired colors together. 
The more experienced artists, on the other hand, said they realized the importance of values even if they did not create value drawings themselves. A1 mentioned that if they had more time, they would have wanted to convert their colored drawing to grayscale to better visualize the values. 
These behaviors illustrate that, although art educational materials might describe a ``canonical'' drawing process, artists have their own adaptations, histories, and workflows. 
Therefore, our tool should not constrain users to a predetermined workflow of drawing.






%% file: 04_tool.tex
\input{tab_system}

\begin{figure*}[t]
    \centering
    \includegraphics[width=\textwidth]{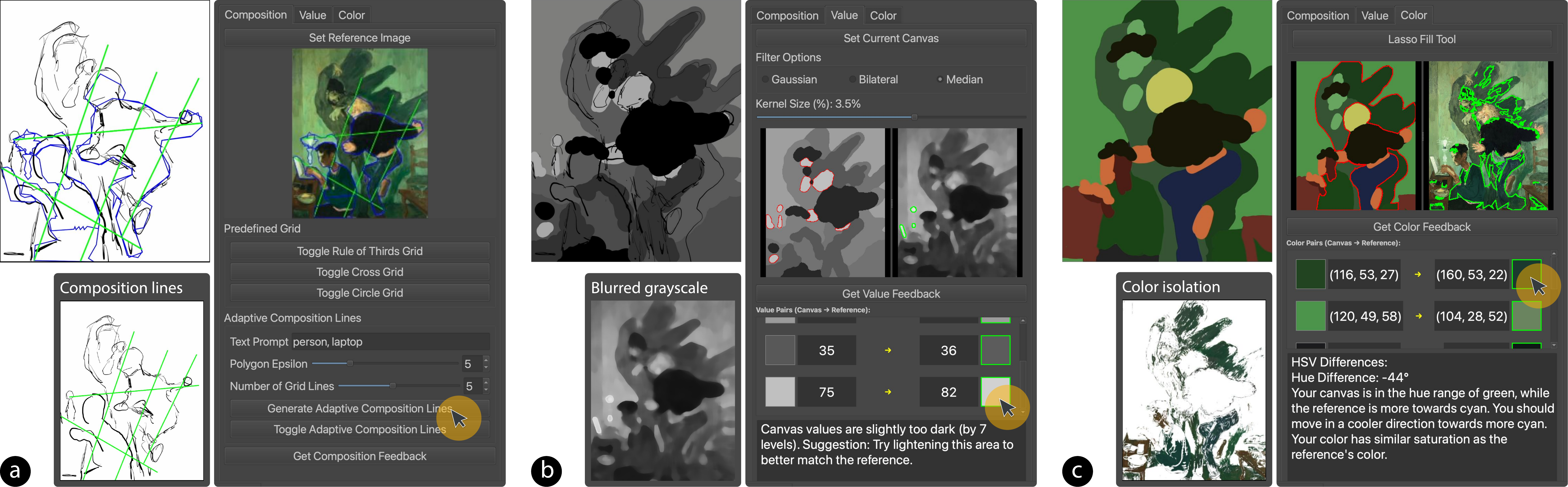}
    \caption{\systemName{} provides guidance (bottom left of each panel) and feedback (right side of each panel, Krita screenshot) for composition (a), value (b), and color (c). Top left of each panel shows the user's progression. 
    We provide guidance and feedback across various modalities, such as overlay, visualization, and verbal descriptions.
    We structure \systemName{} into three tabs to support flexible switching between them.
    \textit{Reference image: The Green Trio by Salman Toor (2019, oil). }
    }
    \label{fig:usage_scenario}
\end{figure*}

\section{\systemName{}}
We introduce the main features of \systemName{} with an example usage scenario (Figure~\ref{fig:usage_scenario}).
We then walk through the automated guidance and feedback that our tool provides for each stage of composition, value, and color (Table~\ref{tab:system}).
We note that users are not required to follow this exact structure of stages and are free to use \systemName{} in any order that aligns with their preferred workflows.

\subsection{Usage scenario}
\label{sec:system_usage}
\textit{Please refer to the video figure which demonstrates this flow.}

\newcommand{\userName}[1]{Arthur}
\newcommand{\userNameTwo}[1]{Frida}

\userName{} wishes to do a study from Salman Toor's work ``The Green Trio'' \cite{Toor2019}.
They open up Krita, where \systemName{} is implemented as a ``docker'' window
and upload the painting as the reference image (Figure~\ref{fig:usage_scenario}).
In the \textit{Composition} tab, they select a rule of thirds grid which the tool overlays on both the reference image and their blank canvas. 
As they start sketching out the two main humans, \userName{} notices they are struggling placing the seated human in the corner.
Therefore, they rectangularly lasso the location of the figure and prompt \systemName{} to look for ``person, laptop.''
\systemName{} then generates adaptive composition lines for an additional visual reference specific to the objects of interest (bottom left of Figure~\ref{fig:usage_scenario}a).
As they finish up their composition draft, they press the button to get feedback, which shows object contours overlaid on their canvas (Figure~\ref{fig:usage_scenario}a). 
Realizing their figure is too large, they edit its size and position before moving on to value.

\userName{} now opens \systemName{}'s \textit{Value} tab, where they reference its blurred black and white thumbnail to take a ``bigger picture'' look at the piece (bottom left of Figure~\ref{fig:usage_scenario}b).
They choose the light, medium, and dark values and start filling in the piece behind the lineart. Once done, they ask \systemName{} for feedback, where they see their values grouped and compared to those of the reference image (Figure~\ref{fig:usage_scenario}b).
This feedback helps \userName{} realize their values need less contrast, so they modify their values 
until satisfied.

Moving on to the \textit{Color} tab, \userName{} uses \systemName{}'s constrained color selection tool to choose just the hue and saturation of colors, while keeping the values constant with what they just painted. They use \systemName{}'s color isolation visualization to notice that the same shade of teal extends across all three figures (bottom left of Figure~\ref{fig:usage_scenario}c). 
After filling in the colors with \systemName{}'s Lasso Fill tool,
\userName{} reads the generated color feedback, presented similar to value feedback (Figure~\ref{fig:usage_scenario}c).
They learn that their choice of the dark green color is similar to the reference in saturation but should be more cyan in hue.
They subsequently use an HSV adjustment layer to make the appropriate edits before shading and rendering the piece (refer to the appendix for the completed drawing).


\subsection{Composition}
\label{sec:system_composition}


\subsubsection{Composition guidance.}
We provide composition guidance in the form of composition lines (Figure~\ref{fig:composition}).
As discussed in Section~\ref{sec:design:findings}, different users prefer various kinds of grids, depending on the reference image content and the stage of their drawing.
Therefore, besides providing user with a set of predefined grids commonly used for composition by artists (rule of thirds~\cite{gooch2001artistic}, central cross, and central circle~\cite{payne1957composition}),
we generate composition lines that describe the spatial relations between objects that the user is currently interested in placing.
%
We based our design decision of using object-based composition on Arnheim's argument that composition arises from ``directed tensions'' of visual elements~\cite{arnheim1954art}.
We support the user to express their intended objects in two ways: (1) describing objects of interests with text and (2) drawing bounding boxes on their canvas (Figure~\ref{fig:composition}).
They can adjust a slider (Figure~\ref{fig:usage_scenario}a) to control how many composition lines they want to see, as the optimal number may vary depending on how many objects they are interested in drawing.

\subsubsection{Composition feedback} 
We display polygon contours of objects in interest directly overlaid on the user's canvas once the user presses the ``Get Composition Feedback'' button (Figure~\ref{fig:usage_scenario}a).
The user can control how coarse the polygon contours are (coarser means less number of edges) by adjusting the polygon epsilon slider (higher value means coarser contours).
We provide polygon approximations, instead of contours closely following the shapes of objects, for two reasons.
The first is that this approximation directly references how artists use the block-in technique to roughly and quickly position objects when drawing~\cite{kolliopoulos2006segmentation}.
The second is that, as revealed in our design study, providing detailed contour might induce the user to directly trace over the lines, instead of practicing how to shape objects themselves~\cite{hertzmann2020line}.

\subsection{Value}
\label{sec:system_value}

\subsubsection{Value guidance}
Humans use their peripheral vision to perceive value (luminance)~\cite{livingstone2002vision}.
We therefore support the user to blur both their canvas and the reference image to mimic peripheral vision, similar to how artists squint their eyes to better see value~\cite{dodson1990keys}.
The user can select three kinds of filters: Gaussian, bilateral, or median, with kernel sizes ranging from 1.5 to 4.9 (Figure~\ref{fig:usage_scenario}b).
Higher kernel size results in more blurring.

\subsubsection{Value feedback} 
To get feedback on value, the user can press the ``Get Value Feedback'' button to see how dominant values in their canvas are compared to those of the reference image.
We display these comparisons as pairs of value swatches.
Clicking on a swatch pair highlights corresponding value regions.
We also provide verbal suggestions on how the user can adjust their value to better match the reference, as motivated by findings in Section~\ref{sec:design:findings}.

\subsection{Color}
\label{sec:system_color}

\subsubsection{Color guidance} 
For color, we provide mechanisms to facilitate color perception and adjustment.
As Albers argues that color perception is dynamic and can be influenced by neighboring colors~\cite{albers2006interaction}, we provide a widget that isolates dominant colors of the reference image as users hover over the region (Figure~\ref{fig:usage_scenario}c).
%
In addition, we provide a widget that applies colors to selected regions on canvas.
The user can use our Lasso Fill feature to select and fill an area with their chosen hue and saturation values.
This can be useful for users moving from the value stage to color their black and white drawings, or for adjusting the existing hue and saturation after reading the verbal color feedback.

\subsubsection{Color feedback}
Feedback for color works similarly to value.
We extract and match dominant colors from the user canvas and the reference image and display color pairs as swatches.
Selecting a pair highlights corresponding regions with verbal feedback on how to adjust hue and saturation (Figure~\ref{fig:usage_scenario}c).

%% file: tab_system.tex
\begin{table}[t]
\caption{For each step of the drawing process, \systemName{} generates corresponding guidance and feedback. Examples can be seen in Figure~\ref{fig:teaser} and \ref{fig:usage_scenario}.
}
\vspace{-2mm}
\resizebox{\linewidth}{!}{%
\begin{tabular}{lll}
\toprule
\textbf{Drawing step} & \textbf{Generated guidance }     & \textbf{Generated feedback} \\ \hline
\vspace {0.5mm}
Composition  & Composition lines       & Polygon contours    \\
\vspace {0.5mm} 
Value        & Blurred grayscale image & Value accuracy     \\
Color        & Color isolation         & Color accuracy \\
\bottomrule
\end{tabular}
}
\label{tab:system}
\end{table}

%% file: 05_method.tex
\section{Implementation}
\label{sec:method}


To enable horizontal movement in existing tool ecosystems \cite{power-cst}, our system is developed as an extension for Krita~\cite{krita}.
We utilize the Krita API~\cite{krita2025api} in Python to interface with Krita and set up a local server in Flask~\cite{flask2025} to send processed data to Krita.

\begin{figure*}[t]
    \centering
    \includegraphics[width=\textwidth]{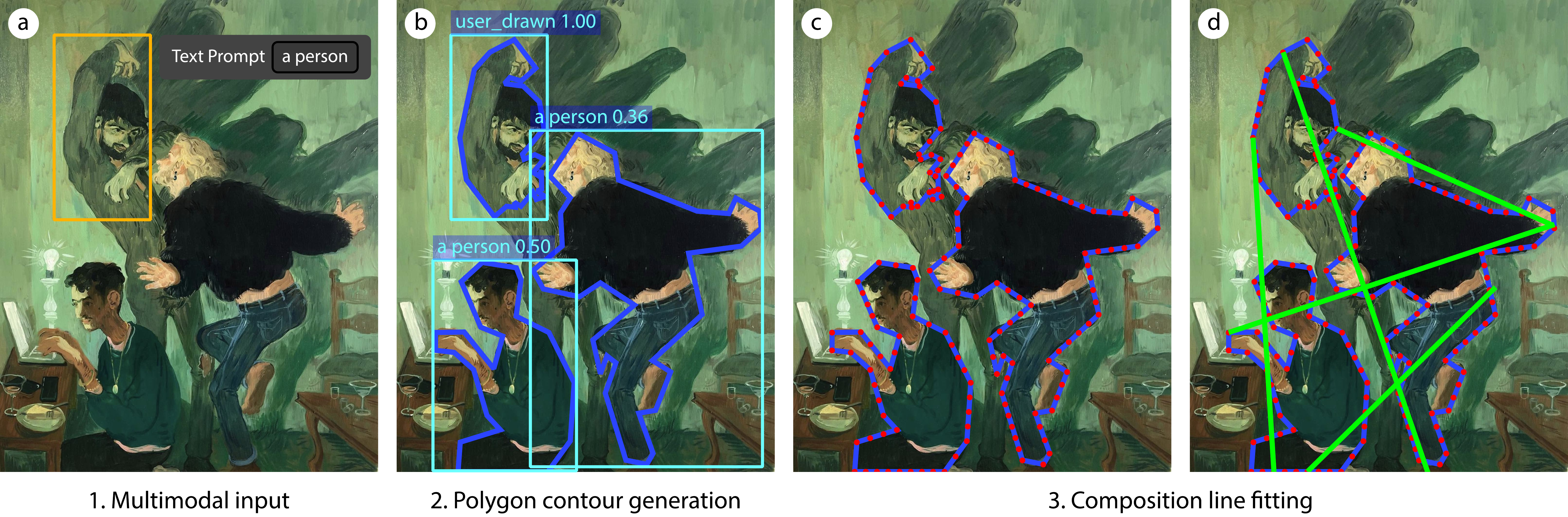}
    \caption{Our pipeline that adaptively generates composition lines supports both textual descriptions and bounding boxes as inputs (a). We first generate polygon contours for objects of interest (b) with Grounded SAM~\cite{ren2024gsam}, and fit composition lines to features points using a modified RANSAC algorithm (c --- d).
    \textit{Reference image: The Green Trio by Salman Toor (2019, oil).}
    }
    \label{fig:composition}
\end{figure*}

\subsection{Adaptive composition line generation}

As shown in Figure~\ref{fig:composition}, our adaptive composition line generation algorithm creates (1) polygon contours of objects of interest as feedback and (2) composition lines that describe the spatial relations between these contours as guidance (Table~\ref{tab:system}).
\changes{On a machine with an NVIDIA GeForce RTX 3090 GPU, the runtime of this algorithm is around two seconds.}

\subsubsection{Polygon contour generation}
Our approach uses the Grounded SAM pipeline~\cite{ren2024gsam} that feeds the bounding boxes detected by Grounding DINO~\cite{liu2024gdino} into Segment Anything (SAM)~\cite{kirillov2023sam} for controlled scene segmentation.
To generate polygon contours, we support two modalities of input: text prompt and vector bounding boxes (Figure~\ref{fig:composition}a).
The user can write in text a list of objects in the reference image that they are interested in, and we directly feed this prompt into Grounding DINO to obtain a set of bounding boxes $\mathcal{B}_{\mathrm{txt}}$.
In addition, the user can also directly draw a set of bounding boxes $\mathcal{B}_{\mathrm{vec}}$ as vector shapes in Krita.
We combine $\mathcal{B}_{\mathrm{vec}}$ and $\mathcal{B}_{\mathrm{txt}}$ into $\mathcal{B}_{\mathrm{all}}$ and send it into SAM to obtain set of segmentation masks $\mathcal{M}_{\mathrm{all}}$.

For each mask $M_{\mathrm{i}} \in \mathcal{M}_{\mathrm{all}}$, we find its outermost contour $C_{\mathrm{i}}$ with the contour approximation algorithm in OpenCV~\cite{opencv2000} without any simplification, so $C_{\mathrm{i}}$ is a smooth contour closely following the shape of an object.
To convert it into a polygon approximation $P_{\mathrm{i}}$, we use the Ramer–Douglas–Peucker algorithm~\cite{douglas1973algorithms} (Figure~\ref{fig:composition}b) and expose the $\varepsilon$ parameter to users in the UI, where a higher $\varepsilon$ produces a coarser approximation (a polygon with fewer edges).

\subsubsection{Composition line fitting.} With a set of polygon contours $P_{\mathrm{i}}$ obtained, we fit lines to delineate the spatial layout formed by these contours by employing the random sample consensus (RANSAC) algorithm~\cite{fischler1981ransac} with several important modifications (Figure~\ref{fig:composition}d). We describe the steps in our algorithm below. Note that we operate in the normalized coordinate system ($[0, 1] \times [0, 1]$).
\begin{enumerate}[leftmargin=1.6em]
    \item Sample points from the polygon contours.
    \item Initialize RANSAC with a line formed by two randomly selected points from \textit{two different polygons}.
    \item Points within an adjustable distance threshold $\theta_\mathrm{dis}=0.04$ from the initial line are considered as inliers. Repeat the line forming process for a fixed number of iterations and record the line with the most amount of inliers as the best fit.
    \item Retain this best fit line as a composition line if it contains inliers more than an adjustable threshold $\theta_\mathrm{inl}$. 
    \item Repeat steps (2) to (4) until no more lines with enough inliers can be found.
\end{enumerate}
%
For step (1), many prior works~\cite{zhou2017ddvp, he2018dtp, ko2024slcd, zhang2023reconstructing} have based this step on lines and edges extracted from images.
However, we find this general approach to be highly sensitive to the quality of lines and edges detected.
This applies more so in our case as changing the $\varepsilon$ parameter results in polygon contours with different sets of edges.
To address this, instead of using polygon edges directly, we sample points from each $P_{\mathrm{i}}$ and uses these as inputs for RANSAC (Figure~\ref{fig:composition}c).
For each edge $E_{\mathrm{i,j}}$ of $P_{\mathrm{i}}$, we first obtain its two endpoints.
As longer edges are likely to be more visually dominant in a composition~\cite{mukherjee2019communicating}, we capture their importance by sampling more points from them.
The number of points sampled is the ratio between the length of edge $E_{\mathrm{i,j}}$ and the length of the shortest edge of $P_{\mathrm{i}}$.
This ensures that points sampled are evenly distributed around the overall contour, with longer edges having more points to influence the RANSAC algorithm later (Figure~\ref{fig:composition}c).

For (2), as composition lines describe relations \textit{between objects}, we initialize the line with points from \textit{different polygons}.
This prevents lines to be fitted to single large polygons with many sampled points.
Finally, for (4), we implement $\theta_\mathrm{inl}$ as a global threshold; each line has to include at least 10\% of all points (determined empirically and adjustable), instead of 10\% of the remaining points.
This ensures that all lines found characterize the global layout structure, instead of small areas of an image.

\begin{figure*}
    \centering
    \includegraphics[width=\textwidth]{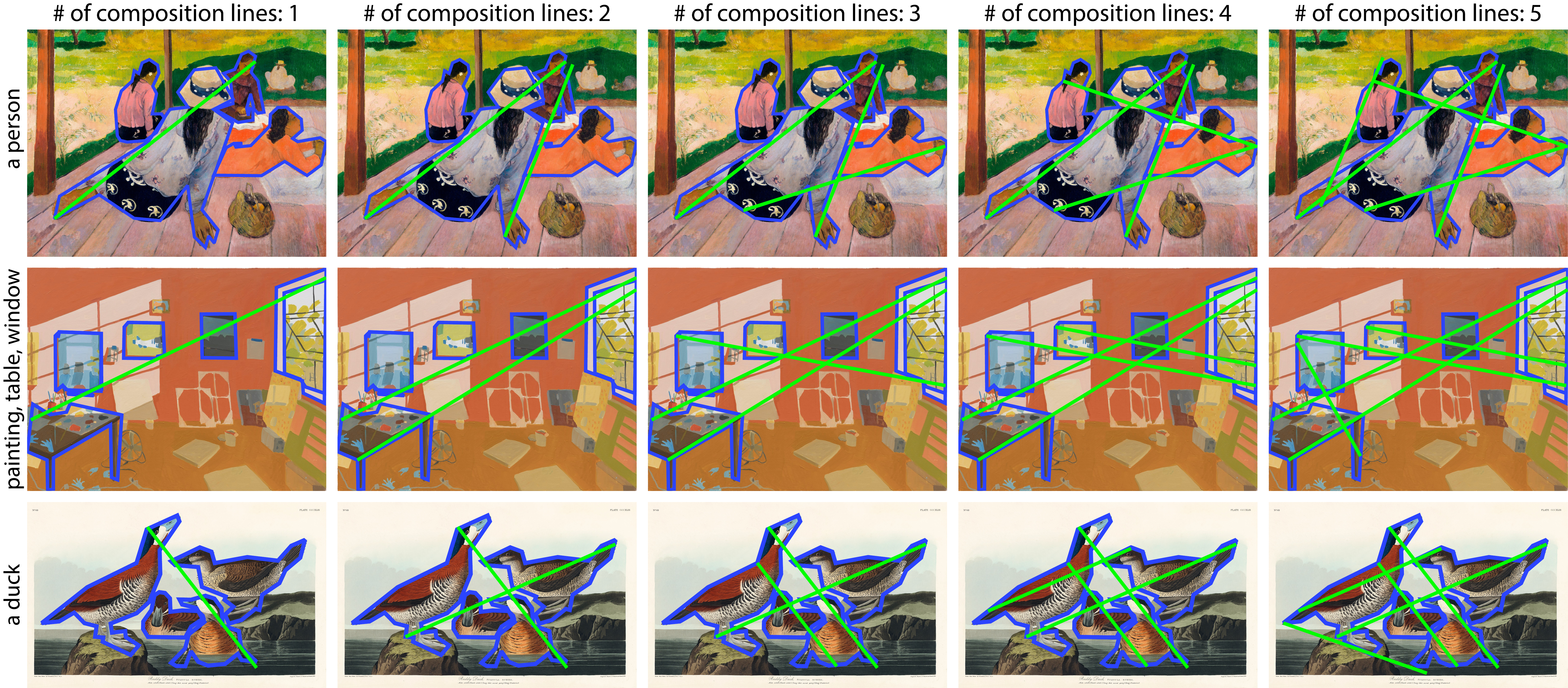}
    \caption{Our pipeline allows user to display the top $k$ composition lines produced via RANSAC fitting.
    While the number of lines vary from image to image, we observe that three to four lines are sufficient to describe prominent spatial relations between objects.
    \textit{Reference images: The Siesta by Paul Gauguin (ca. 1892-94, oil), Red Studio by Sophie Treppendahl (2020, oil), and Ruddy Duck by John James Audubon (1838, watercolor).}
    }
    \label{fig:composition_lines}
\end{figure*}

The sequence in which lines produced by our modified RANSAC naturally encodes their spatial importance in them.
As shown in Figure~\ref{fig:composition_lines}, lines produced earlier cover more points and describe more prominent spatial relations than later lines.
Because of this characteristics, we allow users adjust through the UI to show only the top $k$ lines (Figure~\ref{fig:usage_scenario}a).

\subsection{Value}
To provide feedback on how closely user values match reference image values, we extract dominant values from the user drawn canvas $\mathcal{I}_\mathrm{cnv}$ and the reference image $\mathcal{I}_\mathrm{ref}$ and match them with each other based on spatial and value similarity.
We then display textual descriptions on how each matched pair differ.

\subsubsection{Dominant value extraction}
For an image, we perform value clustering with K-means in the L*a*b* color space, as this representation aligns closer with human visual perception~\cite{mathworks2025lab}.
We filter out extreme values (pure black and white).
For both $\mathcal{I}_\mathrm{cnv}$ and $\mathcal{I}_\mathrm{ref}$, we obtain two sets of dominant values 
$
\mathcal{V}^\mathrm{cnv}$ and $
\mathcal{V}^\mathrm{ref}$.

\subsubsection{Value matching} We match values in $\mathcal{V}^\mathrm{cnv}$ to those in $\mathcal{V}^\mathrm{ref}$ by comparing their similarities in value and spatial locations.
The value similarity between $V^{\mathrm{cnv}}_i \in \mathcal{V}^\mathrm{cnv}$ and $V^{\mathrm{ref}}_j \in \mathcal{V}^\mathrm{ref}$ is given by their normalized Euclidean distance in L*a*b* space~\cite{marccal2023normalised}:
\begin{equation}
    S^{\mathrm{val}}_{i,j} = 1 - \frac{1}{3}\lVert V^{\mathrm{cnv}}_i - V^{\mathrm{ref}}_j \rVert_\mathrm{L*a*b*}
    \label{eq:value}
\end{equation}
We invert this value so that greater is more similar.

To compute the spatial similarity, for a value $V_i$, we create a binary mask $M_i$ where pixels in $\mathcal{I}$ with values within a set threshold of $V_i$ are marked as true.
We empirically determined the threshold to be 5 (L* ranges from 0 to 100).
We then employ OpenCV~\cite{opencv2000} to find contours of all true regions in $M_i$.
We use these contours to visualize regions with the dominant value $V_i$ for the user in the UI.
Given these contours, we find their extreme $x$ and $y$ positions to create a bounding box $B_i$ around them.
We then compute the spatial similarity as:
\begin{equation}
    S^{\mathrm{spt}}_{i,j} = \mathrm{IoU}(B^{\mathrm{cnv}}_i, B^{\mathrm{ref}}_j)
\end{equation}
Putting this together with Equation~\ref{eq:value}, we compute the similarity score between $V^{\mathrm{cnv}}_i$ and $V^{\mathrm{ref}}_j$ as a weighted sum:
\begin{equation}
    S_{i,j} = w_{\mathrm{val}} \cdot S^{\mathrm{val}}_{i,j} + w_{\mathrm{spt}} \cdot S^{\mathrm{spt}}_{i,j}
\end{equation}
We empirically determined that $w_{\mathrm{spt}} = 0.6$ and $w_{\mathrm{val}} = 0.4$.
We build a similarity matrix using $S$ for all values in $\mathcal{V}^\mathrm{cnv}$ and $\mathcal{V}^\mathrm{ref}$ to find a match from $\mathcal{V}^\mathrm{cnv}$ to each value in $\mathcal{V}^\mathrm{spt}$ with the highest score.

\paragraph{Value feedback}
As shown in Figure~\ref{fig:usage_scenario}b, we display the matched value pairs as color swatches. 
Clicking on a swatch highlights corresponding areas on both the canvas and the reference image.
We provide verbal feedback based on the absolute differences between the canvas and reference values.
If the canvas value is darker than the reference, our system would recommend lightening it.
Otherwise, we guide the user to darken it to better match the reference.

\subsection{Color}
As values are grayscale colors, we use the same algorithm as described above to process color.
%
To provide color isolation guidance as described in Section~\ref{sec:system_color} (Table~\ref{tab:system}), we map the color of the pixel hovered over by the user's cursor to one of the dominant colors in $\mathcal{C}^\mathrm{ref}$ and set the rest of the image preview to white to facilitate color perception~\cite{Gurney2010, albers2006interaction}.

\subsubsection{Color feedback}
When a matched pair of colors from $\mathcal{I}_\mathrm{cnv}$ is selected by the user, we compute their difference in hue and saturation to provide verbal feedback (Figure~\ref{fig:usage_scenario}c).
We provide hue feedback on whether the canvas color is warmer or cooler than the reference.
We adopt Photoshop's categorization of hues: red, yellow, green, cyan, blue, and magenta~\cite{photoshop2025hue} and 
map both matched colors into their hue categories.
If the hues are within the same category, our system would recommend a small shift in the warm or cool direction. 
Otherwise, we guide the user to move towards the reference color's hue category. 
For saturation, we compute the difference in percentage to provide feedback on whether the canvas color is brighter or duller compared to the reference color. If the canvas color is brighter, we suggest that the user adjust the color to be less vibrant. Otherwise, we recommend increasing the vibrancy.



%% file: 06_eval.tex
\begin{figure*}[t]
    \centering
    \includegraphics[width=\textwidth]{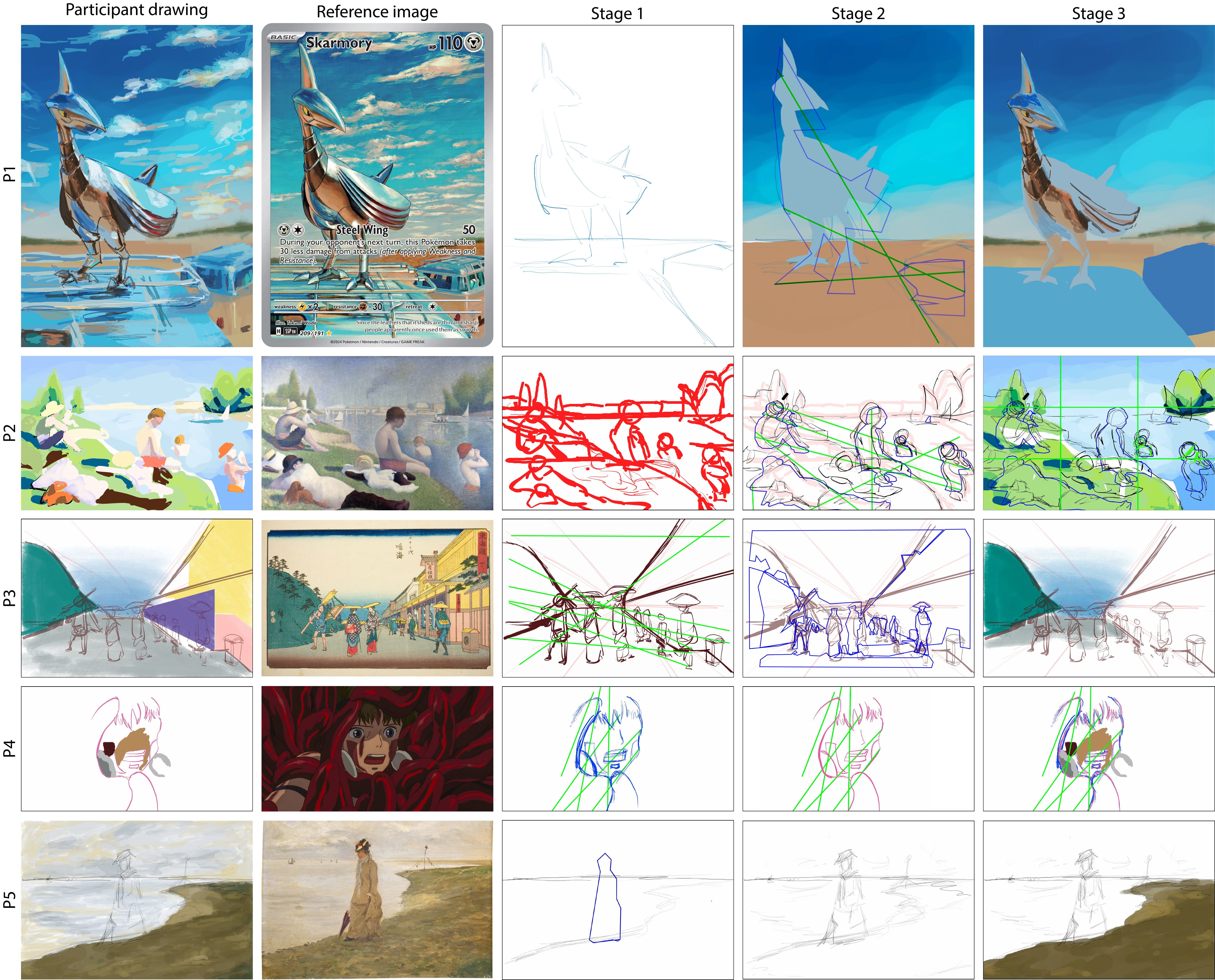}
    \caption{Drawings produced by the participants.
    Notice how the drawing process and the use of \systemName{} vary greatly.
    \textit{Reference images, from top to bottom: Skarmory Pok\'emon card by Takumi Wada (2024, digital), Bathers at Asni\'eres by Georges Seurat (1884, oil), Narumi by Utagawa Hiroshige (1840, woodblock print), still from Princess Mononoke by Studio Ghibli (1997, digital), Woman by the Seaside (artist unknown) (1800s, oil).}
}
    \label{fig:user_evaluation}
\end{figure*}

\section{Evaluation}

In our preliminary evaluation of \systemName{}, we wanted to understand (1) how it fit with existing user workflows, (2) which specific scaffolds were more useful than others, and (3) what new interactions or insights about art arose while using the tool.
%
We conducted two preliminary, IRB-approved evaluations: an in-lab user study with five intermediate artists who used our tool to replicate a reference image of their choice, and gathering baseline feedback on these drawings from two professional artists to compare to \systemName{}'s feedback. In both cases, in addition to validating \systemName{}'s design, we also treated it as a \textit{design probe}~\cite{mattelmaki2006design-probe} that opened larger conversations and insights about drawing practice (reported in Section~\ref{sec:discussion}). 

\subsection{User evaluation procedure}



We conducted a preliminary formative evaluation of \systemName{} with five digital artists. Participants were students at our institution chosen via convenience sampling as long as they self-rated as ``intermediate'' digital artists. All participants had 5-9 years of digital art experience; P1-P4 took traditional drawing courses in school while P5 had no formal art training. The in-lab study took on average two hours and the participants were compensated \$25 USD per hour for their time. As all participants were iPad artists, but Krita is a desktop app, we used Astropad to mirror Krita on the iPad. 

For greater ecological validity, the participants chose their own reference image to replicate. The study consisted of a five-minute pre-interview where we learned about the participants' current approaches to disciplined drawing, a 10-minute tool tutorial (on the \textit{Toor} in Figure~\ref{fig:usage_scenario}), and 90 minutes of free drawing time where the participants were prompted to think aloud and engage (or not engage) with \systemName{}. After drawing, we concluded with a 10-minute post-interview to gather summative feedback on their drawing experience with \systemName{}. We analyzed our data with a reflexive thematic analysis \cite{byrne2022worked}. In line with recent research on creative misuse \cite{creative-misuse}, when we report instances of misuse, it is defined as participants interacting with \systemName{} beyond our initial expectations. 

\subsection{User evaluation findings}

\subsubsection{How did \systemName{} fit in with existing user workflows?}
Each participant had a unique workflow supported by our tool (Figure~\ref{fig:user_evaluation}, Stages 1-3). Although all participants started with a basic sketch, they soon diverged. 
For example, P1 blocked out the dominant colors without any scaffolds before using the composition feedback to determine that the position of the figure was too high.
P4, on the other hand, heavily relied on the composition lines to understand the proportions of the face and refined their sketch into lineart.
P5 misused the polygon contour feedback as an anchor, filling the landscape surrounding the contour first before sketching out the figure. 
While P1 combined color and composition into one step, P2-5 color blocked after they felt like they had a solid composition.


\subsubsection{Reactions to composition, value, and color scaffolds.}
Our participants initially had trouble conceptualizing concrete text prompts to generate composition lines (in line with prior research \cite{johnny-cant-prompt}), such as P3 who prompted the abstract term ``perspective'' before moving to ``a person, a building.'' While the pre-made grids ``lulled [participants] into a false sense of security'' (P2), \systemName{}'s adaptive composition lines were less standardized and normative, with P3 commenting they ``give interesting angular relationships that I probably wouldn't have thought of.'' P4 noted they usually think of facial grids as connecting the eyes, nose, and mouth, so \systemName{}'s lines helped them see in a different way.

All participants heavily used the polygon contour composition feedback to adjust and refine their compositions. They were at a fidelity to help them realize and edit differences in their drawing, retaining autonomy in the interaction. P4 mentioned the feedback prevented simply tracing over the result, while P1 said, ``I got rid of my sketch layer because [the contours] are more accurate, but it's still loose enough that I have to go through and do it myself.''

We observed many instances of misusing the value guidance---none of our participants actually created a grayscale value drawing; P1, P2, and P5 instead used the feature to generate a grayscale version of their canvas to check if their colored rendering and shading had accurate values. P5 commented on this convenience, saying it was ``not something the human eye is capable of doing.'' \systemName{}'s color isolation guided participants to identify \textit{spatial regions} of the same color, such as P5 realizing the browns of the foreground blended with those of the figure. 

Despite value and color feedback having the same interface and interactions, the participants found the value feedback more confusing than color, potentially due to less familiarity with thinking about drawings in grayscale.
In line with prior research \cite{DDB-10.1145/3313831.3376765}, the participants highlighted how the textual feedback helped them interpret the ``scary'' numbers, such as how P2 shifted their greens to be more yellow. P4 interpreted the color feedback as a ``game of how accurately can you interpret a reference,'' and it helped them realize their original colors were too saturated. 

Overall, the usage of certain scaffolds over others was highly dependent on the reference image: for instance, P1 and P5 felt they did not need grids as their image had a simple composition and focused more on color and value, while P2 and P3 had complex compositions and engaged heavily with our scaffolds. 




\subsubsection{How did \systemName{} generate insights and help develop professional vision \cite{goodwin-professional}?}
\systemName{}'s feedback helped the participants develop insights about their current pieces, which they realized also applied to their past drawings.
For example, P2 and P4 noticed their colors were too bright, P2 and P3 noticed their figures were too big, and P1, P3, and P5 noticed their colors were too cool.
Using \systemName{} influenced how participants perceived and reflected on not only their reference images, but also their own workflows. For instance, P2 said using the composition scaffolds forced them to ``slow down'' and deliberately lay down forms. P1 found the grayscale thumbnail so helpful they vowed to start doing value drawings. P4 said the composition lines ``gave you a different way of mapping an image.''


The participants compared \systemName{} to a ``mentor that validates [your drawing], but you have to realize your own mistakes'' (P3) and ``an art teacher [who gives] advice'' (P5). P1 found the scaffolds emotionally helpful, ``I don't do studies because I think they're kind of intimidating, but this does help me help me get over that hurdle.''

\subsubsection{Summary} Overall, \systemName{} adapted to the participants' existing workflows, taught them how to compartmentalize and breakdown a reference image, and encouraged reflection-in-action \cite{schon2017reflective} of both their artwork and process. Through computational guidance and feedback, we found \systemName{} encouraged participants to consciously improve their artistic practices. 




\subsection{Professional evaluation procedure}

We contacted two professional art educators, E1 and E2, to give feedback on drawings made with \systemName{} and \systemName{} itself.
We received feedback on 6 drawings: 4 participant drawings (everyone's but P4's, whose work we thought was too unfinished) and the author created drawings in Figure~\ref{fig:teaser} and \ref{fig:usage_scenario}.
The educators assessed the drawings before being introduced to \systemName{} to avoid biases and to establish an accurate baseline to their teaching methods. Afterwards, to assess \systemName{} itself, the educators saw screen recordings of participant workflows. The study took on average 1 hour and the experts were compensated \$50 USD per hour for their time. 

E1 has been a professional arts educator for 15 years in museum and non-profit contexts and E2 for 30 years in creative electronics and academic contexts. E1's practice focuses on drawing, painting, and printmaking while E2's practice focuses on drawing, 2D design, installation, and screen printing.
We summarize results below; the full professional feedback for each piece is included in the appendix.

\subsection{Professional evaluation findings}
\subsubsection{Feedback overlapping with tool}
Overall, both evaluators gave feedback that focused on composition, value, and color, validating \systemName{}'s breakdown (unsurprising, given we based its design on arts education literature). 
Since all of our user study participants adjusted their composition after seeing \systemName{}'s feedback, experts agreed the general ``shape and positioning of things'' seemed correct. Fewer participants made modifications to value and color so it was subject to critique. For instance, E1 pointed out that P2 should have a darker value on the top left figure's leg to differentiate it from the sand bank, the hands in the dancing figure in Figure \ref{fig:usage_scenario} were too bright, and P5's grass could been more brown, mirroring the insight P5 reported after using the tool (colors too blue).

The educators also commented on participants' processes in a way that aligned with our scaffold design. For example, E2 mentioned instead of sketching the human figure, P5 should have looked at it in terms of color blocks since the figure was more an extension of the landscape---something P5 also realized through \systemName{}'s color isolation. E1 mentioned P3 should have included the long horizontal strips of blue and brown at the top and bottom of the woodblock print---pointing out how the yellow building is overlaid on top of the blue---because those were integral to what made the layered printing process unique. 


\subsubsection{Feedback divergent from tool}
The professional educators also helped identify limitations of \systemName{}---specifically, limitations of properties of drawings that are hard to computationally represent. E1 said they considered all the drawings to be ``value studies,'' focusing on overall composition, value, and color, because they lacked ``drawing'' qualities like rendering the texture of the denim in Figure \ref{fig:usage_scenario}. Mark making properties and stroke quality was a main focus \systemName{} did not support; E1 said the strokes in Figure \ref{fig:teaser}'s reference were much sharper, and that was the main weakness of the study, while E2 disagreed and appreciated the more ``watercolor-like, authentic mannered mark making.'' 

The educators also pointed out different kinds of drawing processes that \systemName{} does not currently support. For instance, E2 said that, when rendering the figures, P3 should have looked at the colors that were visually dominant (i.e., the reds of the kimonos) and captured their gestural impressions, rather than sketching forms as ``character drawing.''

E2 mentioned \systemName{} ``does not show the vibe'' of a composition; they gave an example that illustrating for a bank would result in a heavy-on-bottom composition (to show power and stability), while illustrating for selling skateboards would result in a dynamic and light composition (to show freedom and movement). Composition is tied to visual communication, so \systemName{} could further help intermediate artists see and understand what story their reference image is telling. At the same time, E2 was critical of canonical compositions, saying that visual ``design has been made by middle aged white guys who tell you what the norms are---you have to balance reinforcing those norms with articulating [that the norms are] a starting point.'' 





\subsubsection{Feedback of the tool}
Both educators understood \systemName{}'s purpose as helping participants develop professional vision and reflect on their drawings and processes. ``The whole point is to enable people to learn how to see,'' said E1, while E2 stated, ``It's less about what the tool does and more how you structure people's understanding of it.'' 

In response to our composition line algorithm, E1 stated any line added to an image shows a new spatial relationship between its endpoints. 
E1 went on to explain their own lines for Figure~\ref{fig:composition} would be from the hand of the tall figure to the hand of the dancer, the line connecting the dancer's legs and feet, and a line from the seated figure's head to the dancer's waist, 2/3 of which match our results. E2 interpreted them more as leading lines, saying, ``AI's decisions are so different from ours. It's not representing canonical understanding of the composition, but it analyzes some possible lines of force, even like [prompting users to] think about it.''

With regards to staying in charge of one's process, E1 said, ``[\systemName{}] is not so overly corrective that it's just like copy paste--it's comparing [a user's] artistic interpretation, but it's not correcting it. I like that the tool just shows the people the difference.'' They continued that the tool could ``just make the corrections for you. But that kills the human and the spirit of the process.'' Both educators brought up a tension of concretizing subjective drawing decisions and artistic interpretations, with E1 saying \systemName{} ``sat in the middle...[which] is fun.'' 

Finally, both educators agreed the target user demographic for the tool were illustrators who did not have access to an art teacher. ``It has a great democratizing mission,'' said E1, who additionally pointed out that \systemName{} would be most useful for artists who had the time to slow down, learn, and reflect. They contrasted this with professionals, who were often influenced by other constraints like time, money, and power relationships with clients. 

\subsubsection{Summary} The feedback from two professional art educators was mostly in line with the feedback \systemName{} gave, but highlighted limitations around stroke quality, mark making, and interpretation of artistic intent and norms. Both educators saw the tool as non-restrictive and democratizing learning how to see when drawing.
 





%% file: 07_discussion.tex
\section{Discussion}
\label{sec:discussion}
Through our observations from using \systemName{} as a design probe, we offer speculations and insights for supporting disciplined drawing workflows and computational digital art practices in general, as well as a positioning of why \systemName{} is an artistic support tool. 

\subsection{Reflecting on reflection}

We noticed many instances of reflection-in-action on the artwork encouraged by \systemName{}'s feedback: in changing object positions, lightening values, or shifting color hues.
On the other hand, we also noticed reflection on the user's own processes while engaging with the scaffolds, like how P2 mentioned the composition lines made them ``slow down'' or P1 promising to start doing grayscale value thumbnails.

Recent research has found that feedback-in-action may cause an over-reliance on generated feedback as opposed to engaging in wholistic self-evaluation~\cite{e-feedback-timing}. However, we noticed that despite the frequent timing of asking for feedback, the \textit{coarse} fidelity of \systemName{}'s feedback prevented this over-reliance. For instance, \systemName{}'s object contours for composition feedback were specifically simplified to be rough polygons to avoid tracing. The point of feedback was to let users reflect on their own processes, deeming it ``good enough''---such as when P5 glanced at the extracted color palettes between their canvas and the reference image, ignored the written suggestions, and continued with rendering because they had consciously chosen to feel content with their artwork.


Drawing is a basis for thinking: artists may tediously measure and draw grid lines to think and work through an idea of a composition, more so than to accurately replicate what they see \cite{petherbridge1991primacy}. As \systemName{} generates scaffolds for users, it provides external perceptual support for cognitive tasks. 
However, does using computationally generated scaffolds instead of drafting them manually entirely remove this thinking? We argue no; we still observed reflection from participants responding to the visual information on their canvas. Our purposefully imperfect, ``brittle'' scaffolds served as a starting point for participants to manually think through their art, instead of changing their canvas for them. \changes{This echoes research where participants derived insights from the ``unexpected difference'' between a prompt and a generator's interpretation of the prompt when creating a utopian zine \cite{epstein2022happy} or discovering glitches \cite{prompt-artists-10.1145/3591196.3593515}}. 

The question remains of how (and if) computational scaffolds should nudge users towards greater action and not just realization. For instance, even though P5 realized that the figure and grass merged in color (as E2 also pointed out), they defaulted to their comfortable practice of sketching the whole composition and rendering just the background. Can or should researchers automatically identify the contexts where a nudge may be appropriate to get users to change their workflows for learning?

\subsection{Unanticipated abstraction interpretations}

When designing \systemName{}, we had hoped to fit into existing workflows and justified our argument with teaching materials and a formative study. We were surprised to find several participants saying the \textit{spatial relationships} revealed by color scaffolds, rather than the hue/saturation properties about color, were the most useful. 
We did not anticipate this result.
While we had designed the color scaffolds more for accurate color matching, the participants saw color as no longer a number representing hue, saturation, and value, but something with its own geometric properties. 
Often times the same color spanned multiple distinct objects, so \systemName{}'s color isolation and color feedback visualizations helped participants think of their images in ``color space'' rather than ``object space.'' Computational tools give designers and artists an opportunity to play with these abstractions to reveal new ways of seeing.

\subsection{Digital versus traditional art norms}

Our tool is set in the normative ground \cite{power-cst} of digital art. As mentioned in Section \ref{sec:design:process}, traditional art processes often build up tone and luminance before moving to color. However, none of our participants (who all identified as primarily digital, over traditional, artists) did such a value thumbnail; all jumped straight into color. One reason for this could be that digital artwork is simply a matrix of color values---it lacks the layered and textural elements of traditional art (e.g., the underpainting of an oil paint), so hobbyists who self-teach digital art may not know about value thumbnails since they are not present in the final artifact. When we observed participants ``misusing'' \cite{creative-misuse} \systemName{}'s value feedback to generate a grayscale thumbnail of their current canvas, we saw traditional art norms starting to influence their digital drawing processes. 

Another trend among our participants was that most of their colors were too bright; it was easy to get highly pigmented colors from a digital color picker, but much harder to mix highly pigmented paints and inks for the paintings they were replicating. In addition to viewing this as a manner of technical skill (such as our tool giving feedback that their colors are too bright), what would viewing it as an embrace of the unique affordances and norms of digital art mean? All of our participants chose not to de-saturate their colors. How can HCI researchers build tools that highlight and celebrate the norms of the medium? What would welcoming a chromophilic value system mean in a culture that fears color and treats white as absolute pure and good \cite{batchelor2000chromophobia}? 

Lastly, while our composition lines may not conform to established ``correct'' norms of composition (such as P3 saying they were a ``novel way...not taught in school''), the participants still found them to inspire reflection and new ways of interpreting their reference's composition. Art critic Krauss argues that the turn towards composition grids was art ``turning its back on nature'' to embody the purity only humans were capable of creating~\cite{krauss1979grids}: grids introduced a new norm in illustration practices. Our composition lines are generated with an automated pipeline marrying a traditional algorithm (RANSAC) with artificial intelligence (AI) models (grounded SAM~\cite{ren2024gsam}). By breaking away from traditional grid norms, our composition lines enact new norms influenced by computer algorithms. Beyond practically supporting artistic practices, we argue \systemName{} is an artistic support tool because creating it was an artistic practice to make a value statement: instead of as image generators, why not use AI models as tools to \changes{structure how} artists may see differently? \changes{Rather than de-skilling humans, why not build gyms where artists may exercise their drawing muscles \cite{booten2024build}?}

\subsection{Design suggestions}
From our observations above, we synthesize three design suggestions for future creativity/artistic support tool researchers.
\begin{enumerate}[leftmargin=1.6em]
    \item Computational scaffolds can serve as new materials. When they are \textit{brittle}, they may inspire reflection-in-action without user over-reliance (e.g., rough contour lines prevented tracing). 
    \item When designing ``useful'' scaffolds, researchers should be open to unanticipated encounters that may open new ways of seeing (i.e., users found spatial groups of colors a refreshing departure from object-forward grouping).
    \item Computational scaffolds \changes{that enact new norms} may also drive new insights (e.g., composition lines not grounded in traditional practices created novel visual interpretations). 
\end{enumerate}








\subsection{Limitations and future work}
Several limitations of our tool and study open up directions for future work.
\changes{
While we intended \systemName{} to be domain-agnostic, it might not be suitable for any kind of input image.
Our composition line generation algorithm works best for artworks with defined objects that can be easily described. For more abstract pieces, such as landscapes or De Stijl paintings, users might need to rely on manually drawing bounding boxes to define regions of interest. 
\systemName{}'s color guidance and feedback would provide little value when working with monochrome images, and the value scaffold might not be as constructive for line art drawings.
}

The main limitation of our formative evaluation---despite revealing design insights---was restricted by being in-lab, 2 hours, and a small participant pool unfamiliar with Krita.
\changes{A future longitudinal diary-study with real Krita users will generate an even more meaningful understanding of the ways HCI researchers can support digital drawing, and how CSTs may also serve as socio-technical tools \cite{griffith}.}
Studying \systemName{}'s misuse \cite{creative-misuse} may also reveal additional insights, for instance, how to support non-disciplined drawing. A version of \systemName{} for this in-the-wild deployment would include social learning features, such as the ability to see peers' WIPs, process time-lapses, and feedback, to better emulate the traditional classroom experience.

From our study, we noticed after users laid down the base composition, they often rendered in object-specific regions, such as finishing all the people before moving onto the buildings. To accommodate this workflow, \systemName{} should provide targeted references and feedback on color and value for smaller regions of the artwork.  
Users reflecting on what they learned from \systemName{}'s feedback noted it was in line with mistakes they have historically made (like drawing figures too big). Implementing personalized user profiles would enable \systemName{} to give feedback not just on a single piece, but on trends and artistic growth over time. 


Arnheim, another art critic, argues that a ``good'' composition is one that draws on cultural symbols and reflects the human experience \cite{arnheim1983power}, like the examples E2 gave for laying out a bottom-heavy composition for a bank versus a light and dynamic one for a skateboard shop. This project assumes users want to replicate compositions to learn technical aspects of drawing (e.g., gestalt principles, how objects fit into the rule of thirds). However, we acknowledge that ``technical drawing skills'' cannot be entirely separated from the culture that informs \textit{how} we make drawing decisions in the first place. It is because of culture and human experience that we can interpret compositions in ways that feel meaningful and intriguing, instead of boring and flat (like when objects in a composition perfectly follow a grid and lack tension). We end with a provocation: what would it mean to computationally support a type of disciplined drawing that teaches not just technical skills, but also how to make meaningful cultural commentary through drawing?

%% file: 08_conclusion.tex
\section{Conclusion}
To support digital disciplined drawing, we present \systemName{}, a tool that provides computational feedback and guidance for composition, color, and value. We propose a new algorithm for fitting composition lines that detects objects via Grounded SAM, approximates their contours into polygons, and fits lines to these polygons via RANSAC. \systemName{} adapted to various drawing workflows, helped users realize mismatches between their drawing and the reference image, and encouraged reflection-in-action. Feedback from professional art educators largely aligned with that given by \systemName{} but also accentuated limitations around guiding stroke quality, mark making, and artistic interpretation. Evaluating \systemName{} as a design probe revealed insights for building tools to support illustration: as materials, computational scaffolds should be brittle to prevent over-reliance, and that \changes{analytical} scaffolds and unanticipated interactions can drive new ways of seeing. As an artistic support tool, \systemName{} demonstrates how to integrate prompt-based AI models to serve, rather than displace, existing drawing workflows.

%% file: 00_main.bbl

\begin{thebibliography}{84}


\ifx \showCODEN    \undefined \def \showCODEN     #1{\unskip}     \fi
\ifx \showISBNx    \undefined \def \showISBNx     #1{\unskip}     \fi
\ifx \showISBNxiii \undefined \def \showISBNxiii  #1{\unskip}     \fi
\ifx \showISSN     \undefined \def \showISSN      #1{\unskip}     \fi
\ifx \showLCCN     \undefined \def \showLCCN      #1{\unskip}     \fi
\ifx \shownote     \undefined \def \shownote      #1{#1}          \fi
\ifx \showarticletitle \undefined \def \showarticletitle #1{#1}   \fi
\ifx \showURL      \undefined \def \showURL       {\relax}        \fi
\providecommand\bibfield[2]{#2}
\providecommand\bibinfo[2]{#2}
\providecommand\natexlab[1]{#1}
\providecommand\showeprint[2][]{arXiv:#2}

\bibitem[Adobe(2025)]%
        {photoshop}
\bibfield{author}{\bibinfo{person}{Adobe}.} \bibinfo{year}{2025}\natexlab{}.
\newblock \bibinfo{title}{Photoshop}.
\newblock
\urldef\tempurl%
\url{https://www.adobe.com/products/photoshop.html}
\showURL{%
\tempurl}


\bibitem[Albers and Weber(2006)]%
        {albers2006interaction}
\bibfield{author}{\bibinfo{person}{Josef Albers} {and} \bibinfo{person}{Nicholas~Fox Weber}.} \bibinfo{year}{2006}\natexlab{}.
\newblock \bibinfo{booktitle}{\emph{Interaction of color}}. Vol.~\bibinfo{volume}{6}.
\newblock \bibinfo{publisher}{Yale University Press New Haven}.
\newblock


\bibitem[Arnheim(1954)]%
        {arnheim1954art}
\bibfield{author}{\bibinfo{person}{Rudolf Arnheim}.} \bibinfo{year}{1954}\natexlab{}.
\newblock \bibinfo{booktitle}{\emph{Art and visual perception: A psychology of the creative eye}}.
\newblock \bibinfo{publisher}{Univ of California Press}.
\newblock


\bibitem[Arnheim(1983)]%
        {arnheim1983power}
\bibfield{author}{\bibinfo{person}{Rudolf Arnheim}.} \bibinfo{year}{1983}\natexlab{}.
\newblock \bibinfo{booktitle}{\emph{The power of the center: A study of composition in the visual arts}}.
\newblock \bibinfo{publisher}{Univ of California Press}.
\newblock


\bibitem[Atelier(2025)]%
        {grandatelier2025art}
\bibfield{author}{\bibinfo{person}{Grand~Central Atelier}.} \bibinfo{year}{2025}\natexlab{}.
\newblock \bibinfo{title}{Grand Central Atelier Core Program: Painting Year}.
\newblock \bibinfo{howpublished}{\url{https://grandcentralatelier.org/core-program/painting-year/}}.
\newblock
\newblock
\shownote{Accessed: 2025-03-30}.


\bibitem[Batchelor(2000)]%
        {batchelor2000chromophobia}
\bibfield{author}{\bibinfo{person}{David Batchelor}.} \bibinfo{year}{2000}\natexlab{}.
\newblock \bibinfo{booktitle}{\emph{Chromophobia}}.
\newblock \bibinfo{publisher}{Reaktion books}.
\newblock


\bibitem[Beaudouin-Lafon et~al\mbox{.}(2023)]%
        {color-field}
\bibfield{author}{\bibinfo{person}{Matthew~T Beaudouin-Lafon}, \bibinfo{person}{Jane~L E}, {and} \bibinfo{person}{Haijun Xia}.} \bibinfo{year}{2023}\natexlab{}.
\newblock \showarticletitle{Color Field: Developing Professional Vision by Visualizing the Effects of Color Filters}. In \bibinfo{booktitle}{\emph{Proceedings of the 36th Annual ACM Symposium on User Interface Software and Technology}} (San Francisco, CA, USA) \emph{(\bibinfo{series}{UIST '23})}. \bibinfo{publisher}{Association for Computing Machinery}, \bibinfo{address}{New York, NY, USA}, Article \bibinfo{articleno}{101}, \bibinfo{numpages}{16}~pages.
\newblock
\showISBNx{9798400701320}
\href{https://doi.org/10.1145/3586183.3606828}{doi:\nolinkurl{10.1145/3586183.3606828}}


\bibitem[Booten(2024)]%
        {booten2024build}
\bibfield{author}{\bibinfo{person}{Kyle Booten}.} \bibinfo{year}{2024}\natexlab{}.
\newblock \showarticletitle{Build Word Gyms, Not Word Factories}.
\newblock \bibinfo{journal}{\emph{Critical AI}} \bibinfo{volume}{2}, \bibinfo{number}{1} (\bibinfo{year}{2024}).
\newblock


\bibitem[Bradski(2000)]%
        {opencv2000}
\bibfield{author}{\bibinfo{person}{G. Bradski}.} \bibinfo{year}{2000}\natexlab{}.
\newblock \showarticletitle{{The OpenCV Library}}.
\newblock \bibinfo{journal}{\emph{Dr. Dobb's Journal of Software Tools}} (\bibinfo{year}{2000}).
\newblock


\bibitem[Byrne(2022)]%
        {byrne2022worked}
\bibfield{author}{\bibinfo{person}{David Byrne}.} \bibinfo{year}{2022}\natexlab{}.
\newblock \showarticletitle{A worked example of Braun and Clarke’s approach to reflexive thematic analysis}.
\newblock \bibinfo{journal}{\emph{Quality \& quantity}} \bibinfo{volume}{56}, \bibinfo{number}{3} (\bibinfo{year}{2022}), \bibinfo{pages}{1391--1412}.
\newblock


\bibitem[Calderwood et~al\mbox{.}(2025)]%
        {phraselette-10.1145/3715336.3735832}
\bibfield{author}{\bibinfo{person}{Alex Calderwood}, \bibinfo{person}{John Joon~Young Chung}, \bibinfo{person}{Yuqian Sun}, \bibinfo{person}{Melissa Roemmele}, {and} \bibinfo{person}{Max Kreminski}.} \bibinfo{year}{2025}\natexlab{}.
\newblock \showarticletitle{Phraselette: A Poet’s Procedural Palette}. In \bibinfo{booktitle}{\emph{Proceedings of the 2025 ACM Designing Interactive Systems Conference}} \emph{(\bibinfo{series}{DIS '25})}. \bibinfo{publisher}{Association for Computing Machinery}, \bibinfo{address}{New York, NY, USA}, \bibinfo{pages}{2701–2717}.
\newblock
\showISBNx{9798400714856}
\href{https://doi.org/10.1145/3715336.3735832}{doi:\nolinkurl{10.1145/3715336.3735832}}


\bibitem[Chang et~al\mbox{.}(2023)]%
        {prompt-artists-10.1145/3591196.3593515}
\bibfield{author}{\bibinfo{person}{Minsuk Chang}, \bibinfo{person}{Stefania Druga}, \bibinfo{person}{Alexander~J. Fiannaca}, \bibinfo{person}{Pedro Vergani}, \bibinfo{person}{Chinmay Kulkarni}, \bibinfo{person}{Carrie~J Cai}, {and} \bibinfo{person}{Michael Terry}.} \bibinfo{year}{2023}\natexlab{}.
\newblock \showarticletitle{The Prompt Artists}. In \bibinfo{booktitle}{\emph{Proceedings of the 15th Conference on Creativity and Cognition}} (Virtual Event, USA) \emph{(\bibinfo{series}{C\&C '23})}. \bibinfo{publisher}{Association for Computing Machinery}, \bibinfo{address}{New York, NY, USA}, \bibinfo{pages}{75–87}.
\newblock
\showISBNx{9798400701801}
\href{https://doi.org/10.1145/3591196.3593515}{doi:\nolinkurl{10.1145/3591196.3593515}}


\bibitem[Chernov et~al\mbox{.}(2015)]%
        {chernov2015integer}
\bibfield{author}{\bibinfo{person}{Vladimir Chernov}, \bibinfo{person}{Jarmo Alander}, {and} \bibinfo{person}{Vladimir Bochko}.} \bibinfo{year}{2015}\natexlab{}.
\newblock \showarticletitle{Integer-based accurate conversion between RGB and HSV color spaces}.
\newblock \bibinfo{journal}{\emph{Computers \& Electrical Engineering}}  \bibinfo{volume}{46} (\bibinfo{year}{2015}), \bibinfo{pages}{328--337}.
\newblock


\bibitem[Chevalier et~al\mbox{.}(2012)]%
        {chevalier2012histo}
\bibfield{author}{\bibinfo{person}{Fanny Chevalier}, \bibinfo{person}{Pierre Dragicevic}, {and} \bibinfo{person}{Christophe Hurter}.} \bibinfo{year}{2012}\natexlab{}.
\newblock \showarticletitle{Histomages: fully synchronized views for image editing}. In \bibinfo{booktitle}{\emph{Proceedings of the 25th Annual ACM Symposium on User Interface Software and Technology}} (Cambridge, Massachusetts, USA) \emph{(\bibinfo{series}{UIST '12})}. \bibinfo{publisher}{Association for Computing Machinery}, \bibinfo{address}{New York, NY, USA}, \bibinfo{pages}{281–286}.
\newblock
\showISBNx{9781450315807}
\href{https://doi.org/10.1145/2380116.2380152}{doi:\nolinkurl{10.1145/2380116.2380152}}


\bibitem[Cole et~al\mbox{.}(2008)]%
        {cole2008people}
\bibfield{author}{\bibinfo{person}{Forrester Cole}, \bibinfo{person}{Aleksey Golovinskiy}, \bibinfo{person}{Alex Limpaecher}, \bibinfo{person}{Heather~Stoddart Barros}, \bibinfo{person}{Adam Finkelstein}, \bibinfo{person}{Thomas Funkhouser}, {and} \bibinfo{person}{Szymon Rusinkiewicz}.} \bibinfo{year}{2008}\natexlab{}.
\newblock \showarticletitle{Where do people draw lines?}
\newblock In \bibinfo{booktitle}{\emph{ACM SIGGRAPH 2008 papers}}. \bibinfo{pages}{1--11}.
\newblock


\bibitem[Dixon et~al\mbox{.}(2010)]%
        {dixon2010icandraw}
\bibfield{author}{\bibinfo{person}{Daniel Dixon}, \bibinfo{person}{Manoj Prasad}, {and} \bibinfo{person}{Tracy Hammond}.} \bibinfo{year}{2010}\natexlab{}.
\newblock \showarticletitle{iCanDraw: using sketch recognition and corrective feedback to assist a user in drawing human faces}. In \bibinfo{booktitle}{\emph{Proceedings of the SIGCHI Conference on Human Factors in Computing Systems}} (Atlanta, Georgia, USA) \emph{(\bibinfo{series}{CHI '10})}. \bibinfo{publisher}{Association for Computing Machinery}, \bibinfo{address}{New York, NY, USA}, \bibinfo{pages}{897–906}.
\newblock
\showISBNx{9781605589299}
\href{https://doi.org/10.1145/1753326.1753459}{doi:\nolinkurl{10.1145/1753326.1753459}}


\bibitem[Dodson(1990)]%
        {dodson1990keys}
\bibfield{author}{\bibinfo{person}{Bert Dodson}.} \bibinfo{year}{1990}\natexlab{}.
\newblock \bibinfo{booktitle}{\emph{Keys to drawing}}.
\newblock \bibinfo{publisher}{Penguin}.
\newblock


\bibitem[Douglas and Peucker(1973)]%
        {douglas1973algorithms}
\bibfield{author}{\bibinfo{person}{David~H. Douglas} {and} \bibinfo{person}{Thomas~K. Peucker}.} \bibinfo{year}{1973}\natexlab{}.
\newblock \showarticletitle{Algorithms for the Reduction of the Number of Points Required to Represent a Digitized Line or its Caricature}.
\newblock \bibinfo{journal}{\emph{The Canadian Cartographer}} \bibinfo{volume}{10}, \bibinfo{number}{2} (\bibinfo{year}{1973}), \bibinfo{pages}{112--122}.
\newblock
\href{https://doi.org/10.3138/FM57-6770-U75U-7727}{doi:\nolinkurl{10.3138/FM57-6770-U75U-7727}}


\bibitem[Dow(1914)]%
        {dow1914composition}
\bibfield{author}{\bibinfo{person}{Arthur~Wesley Dow}.} \bibinfo{year}{1914}\natexlab{}.
\newblock \bibinfo{booktitle}{\emph{Composition}}.
\newblock \bibinfo{publisher}{Doubleday, Doran, Incorporated}.
\newblock


\bibitem[Duan et~al\mbox{.}(2024)]%
        {ui-feedback-peitong}
\bibfield{author}{\bibinfo{person}{Peitong Duan}, \bibinfo{person}{Jeremy Warner}, \bibinfo{person}{Yang Li}, {and} \bibinfo{person}{Bjoern Hartmann}.} \bibinfo{year}{2024}\natexlab{}.
\newblock \showarticletitle{Generating Automatic Feedback on UI Mockups with Large Language Models}. In \bibinfo{booktitle}{\emph{Proceedings of the 2024 CHI Conference on Human Factors in Computing Systems}} (Honolulu, HI, USA) \emph{(\bibinfo{series}{CHI '24})}. \bibinfo{publisher}{Association for Computing Machinery}, \bibinfo{address}{New York, NY, USA}, Article \bibinfo{articleno}{6}, \bibinfo{numpages}{20}~pages.
\newblock
\showISBNx{9798400703300}
\href{https://doi.org/10.1145/3613904.3642782}{doi:\nolinkurl{10.1145/3613904.3642782}}


\bibitem[E et~al\mbox{.}(2020)]%
        {e2020composition}
\bibfield{author}{\bibinfo{person}{Jane~L. E}, \bibinfo{person}{Ohad Fried}, \bibinfo{person}{Jingwan Lu}, \bibinfo{person}{Jianming Zhang}, \bibinfo{person}{Radom\'{\i}r M\v{e}ch}, \bibinfo{person}{Jose Echevarria}, \bibinfo{person}{Pat Hanrahan}, {and} \bibinfo{person}{James~A. Landay}.} \bibinfo{year}{2020}\natexlab{}.
\newblock \showarticletitle{Adaptive Photographic Composition Guidance}. In \bibinfo{booktitle}{\emph{Proceedings of the 2020 CHI Conference on Human Factors in Computing Systems}}.
\newblock
\href{https://doi.org/10.1145/3313831.3376635}{doi:\nolinkurl{10.1145/3313831.3376635}}


\bibitem[E et~al\mbox{.}(2024)]%
        {e-feedback-timing}
\bibfield{author}{\bibinfo{person}{Jane~L. E}, \bibinfo{person}{Yu-Chun~Grace Yen}, \bibinfo{person}{Isabelle~Yan Pan}, \bibinfo{person}{Grace Lin}, \bibinfo{person}{Mingyi Li}, \bibinfo{person}{Hyoungwook Jin}, \bibinfo{person}{Mengyi Chen}, \bibinfo{person}{Haijun Xia}, {and} \bibinfo{person}{Steven~P. Dow}.} \bibinfo{year}{2024}\natexlab{}.
\newblock \showarticletitle{When to Give Feedback: Exploring Tradeoffs in the Timing of Design Feedback}. In \bibinfo{booktitle}{\emph{Proceedings of the 16th Conference on Creativity \& Cognition}} (Chicago, IL, USA) \emph{(\bibinfo{series}{C\&C '24})}. \bibinfo{publisher}{Association for Computing Machinery}, \bibinfo{address}{New York, NY, USA}, \bibinfo{pages}{292–310}.
\newblock
\showISBNx{9798400704857}
\href{https://doi.org/10.1145/3635636.3656183}{doi:\nolinkurl{10.1145/3635636.3656183}}


\bibitem[Edwards(1997)]%
        {edwards1997drawing}
\bibfield{author}{\bibinfo{person}{Betty Edwards}.} \bibinfo{year}{1997}\natexlab{}.
\newblock \showarticletitle{Drawing on the Right Side of the Brain}.
\newblock In \bibinfo{booktitle}{\emph{CHI'97 Extended Abstracts on Human Factors in Computing Systems}}. \bibinfo{pages}{188--189}.
\newblock


\bibitem[Epstein et~al\mbox{.}(2022)]%
        {epstein2022happy}
\bibfield{author}{\bibinfo{person}{Ziv Epstein}, \bibinfo{person}{Hope Schroeder}, {and} \bibinfo{person}{Dava Newman}.} \bibinfo{year}{2022}\natexlab{}.
\newblock \showarticletitle{When happy accidents spark creativity: Bringing collaborative speculation to life with generative AI}.
\newblock \bibinfo{journal}{\emph{arXiv preprint arXiv:2206.00533}} (\bibinfo{year}{2022}).
\newblock


\bibitem[Fernquist et~al\mbox{.}(2011)]%
        {fernquist2011ssr}
\bibfield{author}{\bibinfo{person}{Jennifer Fernquist}, \bibinfo{person}{Tovi Grossman}, {and} \bibinfo{person}{George Fitzmaurice}.} \bibinfo{year}{2011}\natexlab{}.
\newblock \showarticletitle{Sketch-sketch revolution: an engaging tutorial system for guided sketching and application learning}. In \bibinfo{booktitle}{\emph{Proceedings of the 24th Annual ACM Symposium on User Interface Software and Technology}} (Santa Barbara, California, USA) \emph{(\bibinfo{series}{UIST '11})}. \bibinfo{publisher}{Association for Computing Machinery}, \bibinfo{address}{New York, NY, USA}, \bibinfo{pages}{373–382}.
\newblock
\showISBNx{9781450307161}
\href{https://doi.org/10.1145/2047196.2047245}{doi:\nolinkurl{10.1145/2047196.2047245}}


\bibitem[Fischler and Bolles(1981)]%
        {fischler1981ransac}
\bibfield{author}{\bibinfo{person}{Martin~A Fischler} {and} \bibinfo{person}{Robert~C Bolles}.} \bibinfo{year}{1981}\natexlab{}.
\newblock \showarticletitle{Random sample consensus: a paradigm for model fitting with applications to image analysis and automated cartography}.
\newblock \bibinfo{journal}{\emph{Commun. ACM}} \bibinfo{volume}{24}, \bibinfo{number}{6} (\bibinfo{year}{1981}), \bibinfo{pages}{381--395}.
\newblock


\bibitem[Foundation(2025)]%
        {krita}
\bibfield{author}{\bibinfo{person}{Krita Foundation}.} \bibinfo{year}{2025}\natexlab{}.
\newblock \bibinfo{title}{Krita (Version 5.2.9)}.
\newblock
\urldef\tempurl%
\url{https://krita.org}
\showURL{%
\tempurl}
\newblock
\shownote{Released on January 29, 2025. Free and open-source digital painting software.}.


\bibitem[Gooch et~al\mbox{.}(2001)]%
        {gooch2001artistic}
\bibfield{author}{\bibinfo{person}{Bruce Gooch}, \bibinfo{person}{Erik Reinhard}, \bibinfo{person}{Chris Moulding}, {and} \bibinfo{person}{Peter Shirley}.} \bibinfo{year}{2001}\natexlab{}.
\newblock \showarticletitle{Artistic composition for image creation}. In \bibinfo{booktitle}{\emph{Rendering Techniques 2001: Proceedings of the Eurographics Workshop in London, United Kingdom, June 25--27, 2001 12}}. Springer, \bibinfo{pages}{83--88}.
\newblock


\bibitem[Goodwin(1994)]%
        {goodwin-professional}
\bibfield{author}{\bibinfo{person}{Charles Goodwin}.} \bibinfo{year}{1994}\natexlab{}.
\newblock \showarticletitle{Professional vision}.
\newblock In \bibinfo{booktitle}{\emph{American Anthropologist}}. \bibinfo{pages}{606--633}.
\newblock


\bibitem[Gurney(2010)]%
        {Gurney2010}
\bibfield{author}{\bibinfo{person}{James Gurney}.} \bibinfo{year}{2010}\natexlab{}.
\newblock \bibinfo{title}{Color Isolator}.
\newblock
\urldef\tempurl%
\url{https://gurneyjourney.blogspot.com/2010/01/color-isolator.html}
\showURL{%
\tempurl}
\newblock
\shownote{Accessed: April 10, 2025}.


\bibitem[Hale(1989)]%
        {hale1989drawing}
\bibfield{author}{\bibinfo{person}{Robert~Beverly Hale}.} \bibinfo{year}{1989}\natexlab{}.
\newblock \bibinfo{booktitle}{\emph{Drawing lessons from the great masters}}.
\newblock \bibinfo{publisher}{Watson-Guptill}.
\newblock


\bibitem[He et~al\mbox{.}(2016)]%
        {he2016deep}
\bibfield{author}{\bibinfo{person}{Kaiming He}, \bibinfo{person}{Xiangyu Zhang}, \bibinfo{person}{Shaoqing Ren}, {and} \bibinfo{person}{Jian Sun}.} \bibinfo{year}{2016}\natexlab{}.
\newblock \showarticletitle{Deep residual learning for image recognition}. In \bibinfo{booktitle}{\emph{Proceedings of the IEEE conference on computer vision and pattern recognition}}. \bibinfo{pages}{770--778}.
\newblock


\bibitem[He et~al\mbox{.}(2018)]%
        {he2018dtp}
\bibfield{author}{\bibinfo{person}{Siqiong He}, \bibinfo{person}{Zihan Zhou}, \bibinfo{person}{Farshid Farhat}, {and} \bibinfo{person}{James~Z. Wang}.} \bibinfo{year}{2018}\natexlab{}.
\newblock \showarticletitle{Discovering Triangles in Portraits for Supporting Photographic Creation}.
\newblock \bibinfo{journal}{\emph{IEEE Transactions on Multimedia}} \bibinfo{volume}{20}, \bibinfo{number}{2} (\bibinfo{year}{2018}), \bibinfo{pages}{496--508}.
\newblock
\href{https://doi.org/10.1109/TMM.2017.2740026}{doi:\nolinkurl{10.1109/TMM.2017.2740026}}


\bibitem[Hennessey et~al\mbox{.}(2017)]%
        {hennessey2017h2s}
\bibfield{author}{\bibinfo{person}{James~W. Hennessey}, \bibinfo{person}{Han Liu}, \bibinfo{person}{Holger Winnemöller}, \bibinfo{person}{Mira Dontcheva}, {and} \bibinfo{person}{Niloy~J. Mitra}.} \bibinfo{year}{2017}\natexlab{}.
\newblock \showarticletitle{How2Sketch: Generating Easy-To-Follow Tutorials for Sketching 3D Objects}.
\newblock \bibinfo{journal}{\emph{Symposium on Interactive 3D Graphics and Games}} (\bibinfo{year}{2017}).
\newblock


\bibitem[Hertzmann(2020)]%
        {hertzmann2020line}
\bibfield{author}{\bibinfo{person}{Aaron Hertzmann}.} \bibinfo{year}{2020}\natexlab{}.
\newblock \showarticletitle{Why do line drawings work? a realism hypothesis}.
\newblock \bibinfo{journal}{\emph{Perception}} \bibinfo{volume}{49}, \bibinfo{number}{4} (\bibinfo{year}{2020}), \bibinfo{pages}{439--451}.
\newblock


\bibitem[Huang et~al\mbox{.}(2022)]%
        {huang2022dualface}
\bibfield{author}{\bibinfo{person}{Zhengyu Huang}, \bibinfo{person}{Yichen Peng}, \bibinfo{person}{Tomohiro Hibino}, \bibinfo{person}{Chunqi Zhao}, \bibinfo{person}{Haoran Xie}, \bibinfo{person}{Tsukasa Fukusato}, {and} \bibinfo{person}{Kazunori Miyata}.} \bibinfo{year}{2022}\natexlab{}.
\newblock \showarticletitle{dualface: Two-stage drawing guidance for freehand portrait sketching}.
\newblock \bibinfo{journal}{\emph{Computational Visual Media}}  \bibinfo{volume}{8} (\bibinfo{year}{2022}), \bibinfo{pages}{63--77}.
\newblock


\bibitem[Iarussi et~al\mbox{.}(2013)]%
        {iarussi2013da}
\bibfield{author}{\bibinfo{person}{Emmanuel Iarussi}, \bibinfo{person}{Adrien Bousseau}, {and} \bibinfo{person}{Theophanis Tsandilas}.} \bibinfo{year}{2013}\natexlab{}.
\newblock \showarticletitle{The drawing assistant: automated drawing guidance and feedback from photographs}. In \bibinfo{booktitle}{\emph{Proceedings of the 26th Annual ACM Symposium on User Interface Software and Technology}} (St. Andrews, Scotland, United Kingdom) \emph{(\bibinfo{series}{UIST '13})}. \bibinfo{publisher}{Association for Computing Machinery}, \bibinfo{address}{New York, NY, USA}, \bibinfo{pages}{183–192}.
\newblock
\showISBNx{9781450322683}
\href{https://doi.org/10.1145/2501988.2501997}{doi:\nolinkurl{10.1145/2501988.2501997}}


\bibitem[Interactive(2025)]%
        {procreate}
\bibfield{author}{\bibinfo{person}{Savage Interactive}.} \bibinfo{year}{2025}\natexlab{}.
\newblock \bibinfo{title}{Procreate}.
\newblock
\urldef\tempurl%
\url{https://procreate.com/}
\showURL{%
\tempurl}


\bibitem[Jiang et~al\mbox{.}(2023)]%
        {ai-art-artists}
\bibfield{author}{\bibinfo{person}{Harry~H. Jiang}, \bibinfo{person}{Lauren Brown}, \bibinfo{person}{Jessica Cheng}, \bibinfo{person}{Mehtab Khan}, \bibinfo{person}{Abhishek Gupta}, \bibinfo{person}{Deja Workman}, \bibinfo{person}{Alex Hanna}, \bibinfo{person}{Johnathan Flowers}, {and} \bibinfo{person}{Timnit Gebru}.} \bibinfo{year}{2023}\natexlab{}.
\newblock \showarticletitle{AI Art and its Impact on Artists}. In \bibinfo{booktitle}{\emph{Proceedings of the 2023 AAAI/ACM Conference on AI, Ethics, and Society}} (Montr\'{e}al, QC, Canada) \emph{(\bibinfo{series}{AIES '23})}. \bibinfo{publisher}{Association for Computing Machinery}, \bibinfo{address}{New York, NY, USA}, \bibinfo{pages}{363–374}.
\newblock
\showISBNx{9798400702310}
\href{https://doi.org/10.1145/3600211.3604681}{doi:\nolinkurl{10.1145/3600211.3604681}}


\bibitem[Kato et~al\mbox{.}(2024)]%
        {griffith}
\bibfield{author}{\bibinfo{person}{Jun Kato}, \bibinfo{person}{Kenta Hara}, {and} \bibinfo{person}{Nao Hirasawa}.} \bibinfo{year}{2024}\natexlab{}.
\newblock \showarticletitle{Griffith: A Storyboarding Tool Designed with Japanese Animation Professionals}. In \bibinfo{booktitle}{\emph{Proceedings of the 2024 CHI Conference on Human Factors in Computing Systems}} (Honolulu, HI, USA) \emph{(\bibinfo{series}{CHI '24})}. \bibinfo{publisher}{Association for Computing Machinery}, \bibinfo{address}{New York, NY, USA}, Article \bibinfo{articleno}{233}, \bibinfo{numpages}{14}~pages.
\newblock
\showISBNx{9798400703300}
\href{https://doi.org/10.1145/3613904.3642121}{doi:\nolinkurl{10.1145/3613904.3642121}}


\bibitem[Kirillov et~al\mbox{.}(2023)]%
        {kirillov2023sam}
\bibfield{author}{\bibinfo{person}{Alexander Kirillov}, \bibinfo{person}{Eric Mintun}, \bibinfo{person}{Nikhila Ravi}, \bibinfo{person}{Hanzi Mao}, \bibinfo{person}{Chloe Rolland}, \bibinfo{person}{Laura Gustafson}, \bibinfo{person}{Tete Xiao}, \bibinfo{person}{Spencer Whitehead}, \bibinfo{person}{Alexander~C. Berg}, \bibinfo{person}{Wan-Yen Lo}, \bibinfo{person}{Piotr Dollar}, {and} \bibinfo{person}{Ross Girshick}.} \bibinfo{year}{2023}\natexlab{}.
\newblock \showarticletitle{Segment Anything}. In \bibinfo{booktitle}{\emph{Proceedings of the IEEE/CVF International Conference on Computer Vision (ICCV)}}. \bibinfo{pages}{4015--4026}.
\newblock


\bibitem[Ko et~al\mbox{.}(2024)]%
        {ko2024slcd}
\bibfield{author}{\bibinfo{person}{Jinwon Ko}, \bibinfo{person}{Dongkwon Jin}, {and} \bibinfo{person}{Chang-Su Kim}.} \bibinfo{year}{2024}\natexlab{}.
\newblock \showarticletitle{Semantic Line Combination Detector}. In \bibinfo{booktitle}{\emph{Proceedings of the IEEE/CVF Conference on Computer Vision and Pattern Recognition (CVPR)}}. \bibinfo{pages}{28066--28075}.
\newblock


\bibitem[Kolliopoulos et~al\mbox{.}(2006)]%
        {kolliopoulos2006segmentation}
\bibfield{author}{\bibinfo{person}{Alexander Kolliopoulos}, \bibinfo{person}{Jack~M Wang}, {and} \bibinfo{person}{Aaron Hertzmann}.} \bibinfo{year}{2006}\natexlab{}.
\newblock \showarticletitle{Segmentation-Based 3D Artistic Rendering.}. In \bibinfo{booktitle}{\emph{Rendering Techniques}}. Citeseer, \bibinfo{pages}{361--370}.
\newblock


\bibitem[Krauss(1979)]%
        {krauss1979grids}
\bibfield{author}{\bibinfo{person}{Rosalind Krauss}.} \bibinfo{year}{1979}\natexlab{}.
\newblock \showarticletitle{Grids}.
\newblock \bibinfo{journal}{\emph{October}}  \bibinfo{volume}{9} (\bibinfo{year}{1979}), \bibinfo{pages}{51--64}.
\newblock


\bibitem[Kreminski and Mateas(2021)]%
        {kreminski2021reflective}
\bibfield{author}{\bibinfo{person}{Max Kreminski} {and} \bibinfo{person}{Michael Mateas}.} \bibinfo{year}{2021}\natexlab{}.
\newblock \showarticletitle{Reflective Creators.}. In \bibinfo{booktitle}{\emph{ICCC}}. \bibinfo{pages}{309--318}.
\newblock


\bibitem[{Krita Development Team}(2025)]%
        {krita2025api}
\bibfield{author}{\bibinfo{person}{{Krita Development Team}}.} \bibinfo{year}{2025}\natexlab{}.
\newblock \bibinfo{booktitle}{\emph{Krita API}}.
\newblock Krita Foundation.
\newblock
\urldef\tempurl%
\url{https://docs.krita.org/en/user_manual/python_scripting.html}
\showURL{%
\tempurl}
\newblock
\shownote{Accessed: 2025-03-28}.


\bibitem[Lee et~al\mbox{.}(2017)]%
        {lee2017sld}
\bibfield{author}{\bibinfo{person}{Jun-Tae Lee}, \bibinfo{person}{Han-Ul Kim}, \bibinfo{person}{Chul Lee}, {and} \bibinfo{person}{Chang-Su Kim}.} \bibinfo{year}{2017}\natexlab{}.
\newblock \showarticletitle{Semantic Line Detection and Its Applications}. In \bibinfo{booktitle}{\emph{Proceedings of the IEEE International Conference on Computer Vision (ICCV)}}.
\newblock


\bibitem[Li et~al\mbox{.}(2025)]%
        {creative-misuse}
\bibfield{author}{\bibinfo{person}{Isabel Li}, \bibinfo{person}{Ace Chen}, \bibinfo{person}{Eric Rawn}, \bibinfo{person}{Shm~Garanganao Almeda}, \bibinfo{person}{Bjoern Hartmann}, {and} \bibinfo{person}{Jingyi. Li}.} \bibinfo{year}{2025}\natexlab{}.
\newblock \showarticletitle{Reimagining Misuse as Creative Practice: Impressions and Implications of Usage Norms on Digital Artists}. In \bibinfo{booktitle}{\emph{Proceedings of the SIGCHI Conference on Human Factors in Computing Systems}} (Yokohama, Japan) \emph{(\bibinfo{series}{CHI '25})}. \bibinfo{publisher}{Association for Computing Machinery}, \bibinfo{address}{New York, NY, USA}.
\newblock
\showISBNx{9798400713941}
\href{https://doi.org/10.1145/3706598.3714068}{doi:\nolinkurl{10.1145/3706598.3714068}}


\bibitem[Li et~al\mbox{.}(2020)]%
        {DDB-10.1145/3313831.3376765}
\bibfield{author}{\bibinfo{person}{Jingyi Li}, \bibinfo{person}{Joel Brandt}, \bibinfo{person}{Radom\'{\i}r Mech}, \bibinfo{person}{Maneesh Agrawala}, {and} \bibinfo{person}{Jennifer Jacobs}.} \bibinfo{year}{2020}\natexlab{}.
\newblock \showarticletitle{Supporting Visual Artists in Programming through Direct Inspection and Control of Program Execution}. In \bibinfo{booktitle}{\emph{Proceedings of the 2020 CHI Conference on Human Factors in Computing Systems}} (Honolulu, HI, USA) \emph{(\bibinfo{series}{CHI '20})}. \bibinfo{publisher}{Association for Computing Machinery}, \bibinfo{address}{New York, NY, USA}, \bibinfo{pages}{1–12}.
\newblock
\showISBNx{9781450367080}
\href{https://doi.org/10.1145/3313831.3376765}{doi:\nolinkurl{10.1145/3313831.3376765}}


\bibitem[Li et~al\mbox{.}(2021)]%
        {learn-visual-artists-li}
\bibfield{author}{\bibinfo{person}{Jingyi Li}, \bibinfo{person}{Sonia Hashim}, {and} \bibinfo{person}{Jennifer Jacobs}.} \bibinfo{year}{2021}\natexlab{}.
\newblock \showarticletitle{What We Can Learn From Visual Artists About Software Development}. In \bibinfo{booktitle}{\emph{Proceedings of the 2021 CHI Conference on Human Factors in Computing Systems}} (Yokohama, Japan) \emph{(\bibinfo{series}{CHI '21})}. \bibinfo{publisher}{Association for Computing Machinery}, \bibinfo{address}{New York, NY, USA}, Article \bibinfo{articleno}{314}, \bibinfo{numpages}{14}~pages.
\newblock
\showISBNx{9781450380966}
\href{https://doi.org/10.1145/3411764.3445682}{doi:\nolinkurl{10.1145/3411764.3445682}}


\bibitem[Li et~al\mbox{.}(2023)]%
        {power-cst}
\bibfield{author}{\bibinfo{person}{Jingyi Li}, \bibinfo{person}{Eric Rawn}, \bibinfo{person}{Jacob Ritchie}, \bibinfo{person}{Jasper Tran~O'Leary}, {and} \bibinfo{person}{Sean Follmer}.} \bibinfo{year}{2023}\natexlab{}.
\newblock \showarticletitle{Beyond the Artifact: Power as a Lens for Creativity Support Tools}. In \bibinfo{booktitle}{\emph{Proceedings of the 36th Annual ACM Symposium on User Interface Software and Technology}} (San Francisco, CA, USA) \emph{(\bibinfo{series}{UIST '23})}. \bibinfo{publisher}{Association for Computing Machinery}, \bibinfo{address}{New York, NY, USA}, Article \bibinfo{articleno}{47}, \bibinfo{numpages}{15}~pages.
\newblock
\showISBNx{9798400701320}
\href{https://doi.org/10.1145/3586183.3606831}{doi:\nolinkurl{10.1145/3586183.3606831}}


\bibitem[Liu et~al\mbox{.}(2024)]%
        {liu2024gdino}
\bibfield{author}{\bibinfo{person}{Shilong Liu}, \bibinfo{person}{Zhaoyang Zeng}, \bibinfo{person}{Tianhe Ren}, \bibinfo{person}{Feng Li}, \bibinfo{person}{Hao Zhang}, \bibinfo{person}{Jie Yang}, \bibinfo{person}{Chun yue Li}, \bibinfo{person}{Jianwei Yang}, \bibinfo{person}{Hang Su}, \bibinfo{person}{Jun-Juan Zhu}, {and} \bibinfo{person}{Lei Zhang}.} \bibinfo{year}{2024}\natexlab{}.
\newblock \showarticletitle{Grounding DINO: Marrying DINO with Grounded Pre-Training for Open-Set Object Detection}. In \bibinfo{booktitle}{\emph{Proceedings of the European Conference on Computer Vision (ECCV)}}.
\newblock


\bibitem[Livingstone and Hubel(2002)]%
        {livingstone2002vision}
\bibfield{author}{\bibinfo{person}{Margaret Livingstone} {and} \bibinfo{person}{David~H Hubel}.} \bibinfo{year}{2002}\natexlab{}.
\newblock \showarticletitle{Vision and art: The biology of seeing}.
\newblock \bibinfo{journal}{\emph{(No Title)}} (\bibinfo{year}{2002}).
\newblock


\bibitem[Loomis(1947)]%
        {loomis1947creative}
\bibfield{author}{\bibinfo{person}{Andrew Loomis}.} \bibinfo{year}{1947}\natexlab{}.
\newblock \bibinfo{booktitle}{\emph{Creative illustration}}.
\newblock \bibinfo{publisher}{Viking Press}.
\newblock


\bibitem[Mar{\c{c}}al(2023)]%
        {marccal2023normalised}
\bibfield{author}{\bibinfo{person}{Andr{\'e}~RS Mar{\c{c}}al}.} \bibinfo{year}{2023}\natexlab{}.
\newblock \showarticletitle{Normalised Color Distances.}. In \bibinfo{booktitle}{\emph{IMPROVE}}. \bibinfo{pages}{134--141}.
\newblock


\bibitem[Mattelm{\"a}ki(2006)]%
        {mattelmaki2006design-probe}
\bibfield{author}{\bibinfo{person}{Tuuli Mattelm{\"a}ki}.} \bibinfo{year}{2006}\natexlab{}.
\newblock \bibinfo{booktitle}{\emph{Design probes}}.
\newblock \bibinfo{publisher}{Aalto University}.
\newblock


\bibitem[Mukherjee et~al\mbox{.}(2019)]%
        {mukherjee2019communicating}
\bibfield{author}{\bibinfo{person}{Kushin Mukherjee}, \bibinfo{person}{Robert~XD Hawkins}, {and} \bibinfo{person}{Judith~E Fan}.} \bibinfo{year}{2019}\natexlab{}.
\newblock \showarticletitle{Communicating semantic part information in drawings}. In \bibinfo{booktitle}{\emph{Proceedings of the Annual Meeting of the Cognitive Science Society}}, Vol.~\bibinfo{volume}{41}.
\newblock


\bibitem[Museum(2025)]%
        {drawing-discipline}
\bibfield{author}{\bibinfo{person}{Princeton University~Art Museum}.} \bibinfo{year}{2025}\natexlab{}.
\newblock \bibinfo{title}{Drawing as Discipline}.
\newblock \bibinfo{howpublished}{Princeton University Art Museum Website}.
\newblock
\urldef\tempurl%
\url{https://artmuseum.princeton.edu/object-package/drawing-discipline/5239}
\showURL{%
\tempurl}


\bibitem[of~Art(2025)]%
        {florenceacademy2025art}
\bibfield{author}{\bibinfo{person}{Florence~Academy of Art}.} \bibinfo{year}{2025}\natexlab{}.
\newblock \bibinfo{title}{Florence Academy of Art Curriculum}.
\newblock \bibinfo{howpublished}{\url{https://www.florenceacademyofart.se/curriculum/}}.
\newblock
\newblock
\shownote{Accessed: 2025-03-30}.


\bibitem[of~Art~Florence(2025)]%
        {angelacademy2025art}
\bibfield{author}{\bibinfo{person}{Angel~Academy of Art~Florence}.} \bibinfo{year}{2025}\natexlab{}.
\newblock \bibinfo{title}{Angel Academy of Art Florence Curriculum}.
\newblock \bibinfo{howpublished}{\url{https://www.artrenewal.org/ateliers/angel-academy-of-art-florence}}.
\newblock
\newblock
\shownote{Accessed: 2025-03-30}.


\bibitem[Payne(1957)]%
        {payne1957composition}
\bibfield{author}{\bibinfo{person}{Edgar~Alwin Payne}.} \bibinfo{year}{1957}\natexlab{}.
\newblock \bibinfo{booktitle}{\emph{Composition of outdoor painting}}.
\newblock \bibinfo{publisher}{EP Payne}.
\newblock


\bibitem[Petherbridge(1991)]%
        {petherbridge1991primacy}
\bibfield{author}{\bibinfo{person}{Deanna Petherbridge}.} \bibinfo{year}{1991}\natexlab{}.
\newblock \bibinfo{booktitle}{\emph{The primacy of drawing: An artist's view}}.
\newblock \bibinfo{publisher}{South Bank Centre}.
\newblock


\bibitem[Photoshop(2025)]%
        {photoshop2025hue}
\bibfield{author}{\bibinfo{person}{Photoshop}.} \bibinfo{year}{2025}\natexlab{}.
\newblock \bibinfo{title}{Apply a Hue/Saturation adjustment}.
\newblock \bibinfo{howpublished}{\url{https://helpx.adobe.com/photoshop/using/adjusting-hue-saturation.html}}.
\newblock
\newblock
\shownote{Accessed: 2025-03-30}.


\bibitem[Qin et~al\mbox{.}(2023)]%
        {qin2023isda}
\bibfield{author}{\bibinfo{person}{Juexiao Qin}, \bibinfo{person}{Xiaohua Sun}, {and} \bibinfo{person}{Weijian Xu}.} \bibinfo{year}{2023}\natexlab{}.
\newblock \showarticletitle{A State-of-Art Review on Intelligent Systems for Drawing Assisting}. In \bibinfo{booktitle}{\emph{Human Interface and the Management of Information}}, \bibfield{editor}{\bibinfo{person}{Hirohiko Mori} {and} \bibinfo{person}{Yumi Asahi}} (Eds.). \bibinfo{publisher}{Springer Nature Switzerland}, \bibinfo{address}{Cham}, \bibinfo{pages}{583--605}.
\newblock
\showISBNx{978-3-031-35132-7}


\bibitem[Ren et~al\mbox{.}(2024)]%
        {ren2024gsam}
\bibfield{author}{\bibinfo{person}{Tianhe Ren}, \bibinfo{person}{Shilong Liu}, \bibinfo{person}{Ailing Zeng}, \bibinfo{person}{Jing Lin}, \bibinfo{person}{Kunchang Li}, \bibinfo{person}{He Cao}, \bibinfo{person}{Jiayu Chen}, \bibinfo{person}{Xinyu Huang}, \bibinfo{person}{Yukang Chen}, \bibinfo{person}{Feng Yan}, \bibinfo{person}{Zhaoyang Zeng}, \bibinfo{person}{Hao Zhang}, \bibinfo{person}{Feng Li}, \bibinfo{person}{Jie Yang}, \bibinfo{person}{Hongyang Li}, \bibinfo{person}{Qing Jiang}, {and} \bibinfo{person}{Lei Zhang}.} \bibinfo{year}{2024}\natexlab{}.
\newblock \bibinfo{title}{Grounded SAM: Assembling Open-World Models for Diverse Visual Tasks}.
\newblock
\showeprint[arxiv]{2401.14159}~[cs.CV]
\urldef\tempurl%
\url{https://arxiv.org/abs/2401.14159}
\showURL{%
\tempurl}


\bibitem[Rockman(2009)]%
        {rockman2009drawing}
\bibfield{author}{\bibinfo{person}{Deborah~A Rockman}.} \bibinfo{year}{2009}\natexlab{}.
\newblock \bibinfo{booktitle}{\emph{Drawing essentials: A guide to drawing from observation}}.
\newblock \bibinfo{publisher}{Oxford University Press}.
\newblock


\bibitem[Ronacher and Project(2025)]%
        {flask2025}
\bibfield{author}{\bibinfo{person}{Armin Ronacher} {and} \bibinfo{person}{Pallets Project}.} \bibinfo{year}{2025}\natexlab{}.
\newblock \bibinfo{booktitle}{\emph{Flask: The Python Microframework for Building Web Applications}}.
\newblock
\urldef\tempurl%
\url{https://palletsprojects.com/p/flask/}
\showURL{%
\tempurl}
\newblock
\shownote{Accessed: 2025-03-28}.


\bibitem[Sch{\"o}n(1983)]%
        {schon2017reflective}
\bibfield{author}{\bibinfo{person}{Donald~A Sch{\"o}n}.} \bibinfo{year}{1983}\natexlab{}.
\newblock \bibinfo{booktitle}{\emph{The reflective practitioner: How professionals think in action}}.
\newblock \bibinfo{publisher}{Routledge}.
\newblock


\bibitem[Schuessler(2025)]%
        {delta-e}
\bibfield{author}{\bibinfo{person}{Zachary Schuessler}.} \bibinfo{year}{2025}\natexlab{}.
\newblock \bibinfo{title}{DeltaE 101}.
\newblock
\urldef\tempurl%
\url{https://zschuessler.github.io/DeltaE/learn/}
\showURL{%
\tempurl}


\bibitem[Shi et~al\mbox{.}(2024)]%
        {shi2024mondrian}
\bibfield{author}{\bibinfo{person}{Xinyu Shi}, \bibinfo{person}{Mingyu Liu}, \bibinfo{person}{Ziqi Zhou}, \bibinfo{person}{Ali Neshati}, \bibinfo{person}{Ryan Rossi}, {and} \bibinfo{person}{Jian Zhao}.} \bibinfo{year}{2024}\natexlab{}.
\newblock \showarticletitle{Exploring Interactive Color Palettes for Abstraction-Driven Exploratory Image Colorization}. In \bibinfo{booktitle}{\emph{Proceedings of the 2024 CHI Conference on Human Factors in Computing Systems}} (Honolulu, HI, USA) \emph{(\bibinfo{series}{CHI '24})}. \bibinfo{publisher}{Association for Computing Machinery}, \bibinfo{address}{New York, NY, USA}, Article \bibinfo{articleno}{146}, \bibinfo{numpages}{16}~pages.
\newblock
\showISBNx{9798400703300}
\href{https://doi.org/10.1145/3613904.3642223}{doi:\nolinkurl{10.1145/3613904.3642223}}


\bibitem[Shugrina et~al\mbox{.}(2020)]%
        {color-triads-10.1145/3386569.3392461}
\bibfield{author}{\bibinfo{person}{Maria Shugrina}, \bibinfo{person}{Amlan Kar}, \bibinfo{person}{Sanja Fidler}, {and} \bibinfo{person}{Karan Singh}.} \bibinfo{year}{2020}\natexlab{}.
\newblock \showarticletitle{Nonlinear color triads for approximation, learning and direct manipulation of color distributions}.
\newblock \bibinfo{journal}{\emph{ACM Trans. Graph.}} \bibinfo{volume}{39}, \bibinfo{number}{4}, Article \bibinfo{articleno}{97} (\bibinfo{date}{Aug.} \bibinfo{year}{2020}), \bibinfo{numpages}{13}~pages.
\newblock
\showISSN{0730-0301}
\href{https://doi.org/10.1145/3386569.3392461}{doi:\nolinkurl{10.1145/3386569.3392461}}


\bibitem[Shugrina et~al\mbox{.}(2017)]%
        {shugrina2017pp}
\bibfield{author}{\bibinfo{person}{Maria Shugrina}, \bibinfo{person}{Jingwan Lu}, {and} \bibinfo{person}{Stephen Diverdi}.} \bibinfo{year}{2017}\natexlab{}.
\newblock \showarticletitle{Playful palette: an interactive parametric color mixer for artists}.
\newblock \bibinfo{journal}{\emph{ACM Trans. Graph.}} \bibinfo{volume}{36}, \bibinfo{number}{4}, Article \bibinfo{articleno}{61} (\bibinfo{date}{July} \bibinfo{year}{2017}), \bibinfo{numpages}{10}~pages.
\newblock
\showISSN{0730-0301}
\href{https://doi.org/10.1145/3072959.3073690}{doi:\nolinkurl{10.1145/3072959.3073690}}


\bibitem[Shugrina et~al\mbox{.}(2019)]%
        {color-builder-10.1145/3290605.3300686}
\bibfield{author}{\bibinfo{person}{Maria Shugrina}, \bibinfo{person}{Wenjia Zhang}, \bibinfo{person}{Fanny Chevalier}, \bibinfo{person}{Sanja Fidler}, {and} \bibinfo{person}{Karan Singh}.} \bibinfo{year}{2019}\natexlab{}.
\newblock \showarticletitle{Color Builder: A Direct Manipulation Interface for Versatile Color Theme Authoring}. In \bibinfo{booktitle}{\emph{Proceedings of the 2019 CHI Conference on Human Factors in Computing Systems}} (Glasgow, Scotland Uk) \emph{(\bibinfo{series}{CHI '19})}. \bibinfo{publisher}{Association for Computing Machinery}, \bibinfo{address}{New York, NY, USA}, \bibinfo{pages}{1–12}.
\newblock
\showISBNx{9781450359702}
\href{https://doi.org/10.1145/3290605.3300686}{doi:\nolinkurl{10.1145/3290605.3300686}}


\bibitem[Sterman et~al\mbox{.}(2022)]%
        {creative-version-contorl-sterman}
\bibfield{author}{\bibinfo{person}{Sarah Sterman}, \bibinfo{person}{Molly~Jane Nicholas}, {and} \bibinfo{person}{Eric Paulos}.} \bibinfo{year}{2022}\natexlab{}.
\newblock \showarticletitle{Towards Creative Version Control}.
\newblock \bibinfo{journal}{\emph{Proc. ACM Hum.-Comput. Interact.}} \bibinfo{volume}{6}, \bibinfo{number}{CSCW2}, Article \bibinfo{articleno}{336} (\bibinfo{date}{Nov.} \bibinfo{year}{2022}), \bibinfo{numpages}{25}~pages.
\newblock
\href{https://doi.org/10.1145/3555756}{doi:\nolinkurl{10.1145/3555756}}


\bibitem[Toor(2019)]%
        {Toor2019}
\bibfield{author}{\bibinfo{person}{Salman Toor}.} \bibinfo{year}{2019}\natexlab{}.
\newblock \bibinfo{title}{The Green Trio}.
\newblock \bibinfo{howpublished}{Painting, oil on panel, 20 x 16 inches}.
\newblock
\urldef\tempurl%
\url{https://marianneboeskygallery.com/exhibitions/28/works/artworks-23904-salman-toor-the-green-trio-2019/}
\showURL{%
\tempurl}
\newblock
\shownote{Exhibited in \textit{Xenia: Crossroads in Portrait Painting}, Marianne Boesky Gallery, New York, Jan 11 – Feb 15, 2020}.


\bibitem[Williford(2017)]%
        {williford2017sketchtivity}
\bibfield{author}{\bibinfo{person}{Blake Williford}.} \bibinfo{year}{2017}\natexlab{}.
\newblock \showarticletitle{SketchTivity: Improving Creativity by Learning Sketching with an Intelligent Tutoring System}. In \bibinfo{booktitle}{\emph{Proceedings of the 2017 ACM SIGCHI Conference on Creativity and Cognition}} (Singapore, Singapore) \emph{(\bibinfo{series}{C\&C '17})}. \bibinfo{publisher}{Association for Computing Machinery}, \bibinfo{address}{New York, NY, USA}, \bibinfo{pages}{477–483}.
\newblock
\showISBNx{9781450344036}
\href{https://doi.org/10.1145/3059454.3078695}{doi:\nolinkurl{10.1145/3059454.3078695}}


\bibitem[Williford et~al\mbox{.}(2019)]%
        {williform2019drawmyphoto}
\bibfield{author}{\bibinfo{person}{Blake Williford}, \bibinfo{person}{Abhay Doke}, \bibinfo{person}{Michel Pahud}, \bibinfo{person}{Ken Hinckley}, {and} \bibinfo{person}{Tracy Hammond}.} \bibinfo{year}{2019}\natexlab{}.
\newblock \showarticletitle{DrawMyPhoto: Assisting Novices in Drawing from Photographs}. In \bibinfo{booktitle}{\emph{Proceedings of the 2019 Conference on Creativity and Cognition}} (San Diego, CA, USA) \emph{(\bibinfo{series}{C\&C '19})}. \bibinfo{publisher}{Association for Computing Machinery}, \bibinfo{address}{New York, NY, USA}, \bibinfo{pages}{198–209}.
\newblock
\showISBNx{9781450359177}
\href{https://doi.org/10.1145/3325480.3325507}{doi:\nolinkurl{10.1145/3325480.3325507}}


\bibitem[Works(2025)]%
        {mathworks2025lab}
\bibfield{author}{\bibinfo{person}{Math Works}.} \bibinfo{year}{2025}\natexlab{}.
\newblock \bibinfo{title}{Color-Based Segmentation Using K-Means Clustering}.
\newblock \bibinfo{howpublished}{\url{https://www.mathworks.com/help/images/color-based-segmentation-using-k-means-clustering.html}}.
\newblock
\newblock
\shownote{Accessed: 2025-03-30}.


\bibitem[Xie et~al\mbox{.}(2014)]%
        {xie2014ps}
\bibfield{author}{\bibinfo{person}{Jun Xie}, \bibinfo{person}{Aaron Hertzmann}, \bibinfo{person}{Wilmot Li}, {and} \bibinfo{person}{Holger Winnem\"{o}ller}.} \bibinfo{year}{2014}\natexlab{}.
\newblock \showarticletitle{PortraitSketch: face sketching assistance for novices}. In \bibinfo{booktitle}{\emph{Proceedings of the 27th Annual ACM Symposium on User Interface Software and Technology}} (Honolulu, Hawaii, USA) \emph{(\bibinfo{series}{UIST '14})}. \bibinfo{publisher}{Association for Computing Machinery}, \bibinfo{address}{New York, NY, USA}, \bibinfo{pages}{407–417}.
\newblock
\showISBNx{9781450330695}
\href{https://doi.org/10.1145/2642918.2647399}{doi:\nolinkurl{10.1145/2642918.2647399}}


\bibitem[Zamfirescu-Pereira et~al\mbox{.}(2023)]%
        {johnny-cant-prompt}
\bibfield{author}{\bibinfo{person}{J.D. Zamfirescu-Pereira}, \bibinfo{person}{Richmond~Y. Wong}, \bibinfo{person}{Bjoern Hartmann}, {and} \bibinfo{person}{Qian Yang}.} \bibinfo{year}{2023}\natexlab{}.
\newblock \showarticletitle{Why Johnny Can’t Prompt: How Non-AI Experts Try (and Fail) to Design LLM Prompts}. In \bibinfo{booktitle}{\emph{Proceedings of the 2023 CHI Conference on Human Factors in Computing Systems}} (Hamburg, Germany) \emph{(\bibinfo{series}{CHI '23})}. \bibinfo{publisher}{Association for Computing Machinery}, \bibinfo{address}{New York, NY, USA}, Article \bibinfo{articleno}{437}, \bibinfo{numpages}{21}~pages.
\newblock
\showISBNx{9781450394215}
\href{https://doi.org/10.1145/3544548.3581388}{doi:\nolinkurl{10.1145/3544548.3581388}}


\bibitem[Zhang et~al\mbox{.}(2023b)]%
        {zhang2023reconstructing}
\bibfield{author}{\bibinfo{person}{Jing Zhang}, \bibinfo{person}{R{\'e}mi Synave}, \bibinfo{person}{Samuel Delepoulle}, {and} \bibinfo{person}{R{\'e}mi Cozot}.} \bibinfo{year}{2023}\natexlab{b}.
\newblock \showarticletitle{Reconstructing Image Composition: Computation of Leading Lines}.
\newblock \bibinfo{journal}{\emph{Journal of Imaging}} \bibinfo{volume}{10}, \bibinfo{number}{1} (\bibinfo{year}{2023}), \bibinfo{pages}{5}.
\newblock


\bibitem[Zhang et~al\mbox{.}(2023a)]%
        {zhang2023controlnet}
\bibfield{author}{\bibinfo{person}{Lvmin Zhang}, \bibinfo{person}{Anyi Rao}, {and} \bibinfo{person}{Maneesh Agrawala}.} \bibinfo{year}{2023}\natexlab{a}.
\newblock \showarticletitle{Adding Conditional Control to Text-to-Image Diffusion Models}. In \bibinfo{booktitle}{\emph{Proceedings of the IEEE/CVF International Conference on Computer Vision (ICCV)}}. \bibinfo{pages}{3836--3847}.
\newblock


\bibitem[Zhao et~al\mbox{.}(2022)]%
        {zhao2022hough}
\bibfield{author}{\bibinfo{person}{Kai Zhao}, \bibinfo{person}{Qi Han}, \bibinfo{person}{Chang-Bin Zhang}, \bibinfo{person}{Jun Xu}, {and} \bibinfo{person}{Ming-Ming Cheng}.} \bibinfo{year}{2022}\natexlab{}.
\newblock \showarticletitle{Deep Hough Transform for Semantic Line Detection}.
\newblock \bibinfo{journal}{\emph{IEEE Transactions on Pattern Analysis and Machine Intelligence}} \bibinfo{volume}{44}, \bibinfo{number}{9} (\bibinfo{year}{2022}), \bibinfo{pages}{4793--4806}.
\newblock
\href{https://doi.org/10.1109/TPAMI.2021.3077129}{doi:\nolinkurl{10.1109/TPAMI.2021.3077129}}


\bibitem[Zhou et~al\mbox{.}(2017)]%
        {zhou2017ddvp}
\bibfield{author}{\bibinfo{person}{Zihan Zhou}, \bibinfo{person}{Farshid Farhat}, {and} \bibinfo{person}{James~Z. Wang}.} \bibinfo{year}{2017}\natexlab{}.
\newblock \showarticletitle{Detecting Dominant Vanishing Points in Natural Scenes with Application to Composition-Sensitive Image Retrieval}.
\newblock \bibinfo{journal}{\emph{IEEE Transactions on Multimedia}} \bibinfo{volume}{19}, \bibinfo{number}{12} (\bibinfo{year}{2017}), \bibinfo{pages}{2651--2665}.
\newblock
\href{https://doi.org/10.1109/TMM.2017.2703954}{doi:\nolinkurl{10.1109/TMM.2017.2703954}}


\end{thebibliography}
